\pdfoutput=1
\documentclass[a4paper,fleqn,usenatbib]{mnras}
\usepackage[T1]{fontenc}
\usepackage{ae,aecompl}

\usepackage{graphicx}	% Including figure files
\usepackage{amsmath}	% Advanced maths commands
\usepackage{amssymb}	% Extra maths symbols

\usepackage{enumerate}
\usepackage{longtable}
\usepackage {threeparttable}
\usepackage{lipsum}
\usepackage{xcolor}
\usepackage{url}
\usepackage{newtxtext,newtxmath}
\title[Thermodynamic properties of small flares observed by H$\alpha$ and EUV]{Thermodynamic properties of small flares in the quiet Sun observed by H$\alpha$ and EUV: plasma motion of the chromosphere and time evolution of temperature / emission measure}

\author[Y. Kotani et al.]{
Yuji Kotani,$^{1}$\thanks{E-mail: kotani@kusastro.kyoto-u.ac.jp}
T. T. Ishii,$^{1}$
D. Yamasaki,$^{1, 4}$
K. Otsuji,$^{2}$
K. Ichimoto,$^{1}$
A. Asai,$^{1}$
and K. Shibata$^{3}$
\\
$^{1}$Astronomical Observatory, Kyoto University, Sakyo, Kyoto 606-8502, Japan\\
$^{2}$Space Environment Laboratory, Applied Electromagnetic Research Institute,\\
National Institute of Information and Communications Technology, Koganei, Tokyo 184-8795, Japan\\
$^{3}$Department of Environmental Systems Science, Faculty of Science and Engineering, Doshisha University,\\ 1-3, Tatara Miyakodani, Kyotanabe City, Kyoto, 610-0394, Japan\\
$^{4}$Institute of Space and Astronautical Science, Japan Aerospace Exploration Agency, 3-1-1, Yoshinodai, Chuo-ku, Sagamihara, Kanagawa, 252-5210, Japan
}

\date{Accepted XXX. Received YYY; in original form ZZZ}

\pubyear{2022}

   \volume{{\rm in press}}
\begin{document}
\label{firstpage}
\pagerange{\pageref{firstpage}--\pageref{lastpage}}
\maketitle

\begin{abstract}
Small flares frequently occur in the quiet Sun.
Previous studies have noted that they share many common characteristics with typical solar flares in active regions.
However, their similarities and differences are not fully understood, especially their thermal properties.
In this study, we performed imaging spectroscopic observations in the H$\alpha$ line taken with the \textit{Solar Dynamics Doppler Imager} on the \textit{Solar Magnetic Activity Research Telescope} (SMART/SDDI) at the Hida Observatory and imaging observations with the \textit{Atmospheric Imaging Assembly} onboard \textit{Solar Dynamics Observatory} (SDO/AIA).
We analysed 25 cases of small flares in the quiet Sun over the thermal energy range of $10^{24}-10^{27}\,\mathrm{erg}$, paying particular attention to their thermal properties. Our main results are as follows:
(1) We observe a redshift together with line centre brightening in the H$\alpha$ line associated with more than half of the small flares.
(2) We employ differential emission measure analysis using AIA multi-temperature (channel) observations to obtain the emission measure and temperature of the small flares.
The results are consistent with the \citet{1999ApJ...526L..49S,2002ApJ...577..422S} scaling law. %Shibata \& Yokoyama scaling law.
From the scaling law, we estimated the coronal magnetic field strength of small flares to be 5 --15 G.
(3) The temporal evolution of the temperature and the density shows that the temperature peaks precede the density peaks in more than half of the events.
These results suggest that chromospheric evaporations/condensations play an essential role in the thermal properties of  some of the small flares in the quiet Sun, as does for large flares.

\end{abstract}


%%%%%%%%%%%%%%%%%%%%%%%%%%%%%%%%%%%%%%%%%%%%%%%%%%

%%%%%%%%%%%%%%%%% BODY OF PAPER %%%%%%%%%%%%%%%%%%

\section{Introduction}     \label{sec1}

Small flares are frequently observed throughout the solar atmosphere.
Depending on their energy, they have been named microflares ($10^{26}-10^{29}\,\mathrm{erg}$) or nanoflares ($10^{23}-10^{26}\,\mathrm{erg}$).
In this paper, we focus on small flares in the quiet Sun (QS) corona with energies of  $10^{24}-10^{27}\,\mathrm{erg}$ (hereafter ``small flares'').
%The temperature of these brightenings is $<2\times10^6\,\mathrm{K}$ \citep[e.g.,][]{2000ApJ...535.1047A}, smaller than the small flares in the active region with comparable energies.

%Small flares in the quiet region (QR) in the solar corona with energies of $10^{24}-10^{26}\,\mathrm{erg}$ are frequently observed.
%The temperature of these brightenings is $<2\times10^6\,\mathrm{K}$ \citep[e.g.,][]{2000ApJ...535.1047A}, smaller than the small flares in the active region with comparable energies.
Various instruments have observed brightenings associated with small flares for more than 20 years \citep{1997ApJ...488..499K,1998ApJ...501L.213K,1998A&A...336.1039B,2000ApJ...535.1047A,2000ApJ...529..554P,2002ApJ...568..413B,2016A&A...591A.148J,2021A&A...647A.159C,2022A&A...661A.149P}.
%Many previous studies have investigated whether these small brightenings have the required energy to heat the corona  \citep[nanoflare heating problem, ][]{1988ApJ...330..474P}.
Numerous studies have investigated whether small flares have the required energies to heat the solar corona \citep{1988ApJ...330..474P}.
It is believed that the nanoflares observed with the current telescope resolution do not have enough energy to heat the steady corona \citep{2021A&A...647A.159C}.
In contrast, we have not yet reached a consensus on whether nanoflares that are too small to be captured by the current resolution can be responsible for coronal heating.
For nanoflares to have a significant contribution to coronal heating, the power-law index between the flare frequency $dN/dE$ and the flare energy $E$ must be smaller than $-2$ ($dN/dE\propto E^{\alpha}$, $\alpha<-2$) \citep{1991SoPh..133..357H}.
In recent observations using the \textit{Atmospheric Imaging Assembly} onboard the \textit{Solar Dynamics Observatory} (SDO/AIA) \citep{2012SoPh..275....3P,2012SoPh..275...17L}, \citet{2022A&A...661A.149P} reported a power-law index smaller than $-2$ ($\alpha=-2.28\pm0.03$), whilst \citet{2016A&A...591A.148J} reported a power-law index larger than $-2$ ($\alpha=-1.73$ in the QS).
Further understanding of the mechanism of small flares in the QS is crucially important for resolving this open question.

Small flares in the QS at coronal temperatures often accompany small eruptions at chromospheric temperatures (called ``minifilament"). 
Minifilament eruptions were first reported in the 1970s \citep{1977ApJ...218..286M,1979SoPh...61..283L}. 
Multi-wavelength observations using chromospheric and Extreme Ultraviolet (EUV) lines in the 2000s showed that they typically accompany coronal small flares \citep{2004ApJ...616..578S,2008Ap&SS.318..141R}.
These studies also reported that minifilament eruptions were associated with magnetic flux cancellation in the photosphere. 
Minifilament eruptions are often reported to be associated with jets \citep[e.g.,][]{2015Natur.523..437S}.
\citet{2020A&A...643A..19M} reported simultaneous brightening in the H$\alpha$ line and EUV accompanied by a minfilament eruption.
\citet{2019A&A...623A..78G} reproduced the coronal magnetic field  at the location of a microflare and minifilament eruption in a coronal bright point \citep[CBP, ][]{2019LRSP...16....2M} using the non-linear force-free field (NLFFF) method.
They found that twisted magnetic  field lines (flux ropes) form at the minifilament eruption location.
This is similar to the magnetic morphology of solar flares \citep[e.g.,][]{2013ApJ...771L..30J,2014ApJ...788..182I,2022arXiv221014563Y}.
\citet{2020ApJ...898..144K} and \citet{2022ApJ...939...25P} performed spectroscopic observations of chromospheric lines (H$\alpha$ and Mg II, respectively) and reported that brightenings and downflow in the chromosphere are observed in response to EUV brightenings.
\citet{2021ApJ...914L..35J} observed microflares with minifilament eruptions using the AIA. 
They found that the AIA 304 \AA\, light curve's peak precedes the coronal emission's peak by 2 or 3 minutes.
This trend suggests a Neupert effect in large flares \citep{1968ApJ...153L..59N}; that is, the hard X-ray light curve corresponds to the time derivative of the soft X-ray light curve.
All these properties support the interpretation that small flares in the QS with minifilament eruptions are miniature versions of typical solar flares associated with filament eruptions.

Recent observations by the Extreme Ultraviolet Imager \citep[EUI,][]{2020A&A...642A...8R} onboard Solar Orbiter \citep[SO,][]{2020A&A...642A...1M} revealed  small brightenings in the QS that were named ``campfires'' \citep{2021A&A...656L...4B}.
The average temperature of these brightenings was estimated at $\log T=6.1$, which is consistent with other brightenings observed in the QS in the past.
\citet{2021A&A...656L...4B} and \citet{2021A&A...656A..35Z} compared their length and height using triangulation with SO/EUI and SDO/AIA.
These studies found that the height was larger than the length, which suggests that the brightenings occur only near the apex of the loops.
\citet{2021ApJ...921L..20P} found that campfires are often associated with magnetic flux cancellation and dark eruptions seen in EUV emission that may correspond to minifilament.
With recent improvements in numerical simulation techniques, realistic simulations have been performed to reproduce the EUI brightenings \citep[e.g.,][]{2022ApJ...929..103T,2023ApJ...943...24P}.
\citet{2021A&A...656L...7C} performed the 3D radiation magnetohydrodynamics simulation with a QS parameter and reproduced brightenings with similar properties to the observed features.
They found that these brightenings are caused by heating the cool and dense plasma to 1 MK. 
This heating is due to magnetic reconnection that occurs below the transition region. 
This mechanism differs from the brightening mechanism involving chromospheric evaporation when reconnection occurs in the corona.

Although small flares in the QS have been studied intensively, the similarities and differences with the physical mechanisms of typical solar flares are not fully understood.
In particular, we do not fully understand how small-scale reconnection events affect the chromosphere.
Several minifilament eruption studies have reported brightening in the chromosphere with EUV brightening \citep[e.g.,][]{2004ApJ...616..578S,2020A&A...643A..19M}.
However, only \citet{2020ApJ...898..144K}, \citet{2022A&A...660A..45M}, and \citet{2022ApJ...939...25P} performed spectroscopic observations, and all studies analysed only one event.
Investigating whether small flares in the QS also show red asymmetry in chromospheric lines observed during large flares \citep{1984SoPh...93..105I, 1990ApJ...363..318C} may help us understand the heating mechanism at work \citep{2021ApJ...912...25A}.
To the best of our knowledge, there exist no studies clearly showing chromospheric evaporation at work during small flares in the QS.
A comparison with scaling law studies of flares in active regions is also expected to help investigate the physical mechanism of small flares in the QS \citep[e.g.,][]{1999ApJ...526L..49S,2002ApJ...577..422S,2017ApJ...851...91N,2020ApJ...903...23A}.
If the observations are consistent with the scaling law, we can expect the same physical mechanisms at work in small flares in the QS as in large flares.

In this study, we performed H$\alpha$ line imaging and spectroscopic analysis together with coronal EUV imaging analysis of more than 20 small flares in the QS.
We aimed to improve our understanding of the thermal evolution of small flares in the QS and their impact on the chromosphere.
This paper uses the term "small flare" while referring to brightening phenomena in EUV emission at coronal temperature.

\section{Observations and data processing}      \label{sec2}

We observed the chromosphere using the \textit{Solar Dynamics Doppler Imager} \citep[SDDI: ][]{2017SoPh..292...63I} on the \textit{Solar Magnetic Activity Research Telescope} \citep[SMART: ][]{2004SPIE.5492..958U} at the Hida Observatory of Kyoto University.
SDDI performs H$\alpha$ imaging spectroscopy observations of the full solar disk with a time resolution of 12 s and a pixel size of 1.23 arcsec.
In this study, we took H$\alpha$ images at 25 wavelengths from H$\alpha -3.0$ \AA\, to H$\alpha +3.0$ \AA\, with a constant wavelength step of 0.25 \AA.
The data were processed with dark and flat field corrections.
We used a position angle for the Sun to rotate the SDDI data and make the upward direction of the image orientate towards the solar north pole.
We confirmed that the rotated SDDI data and AIA level 1.5 data generated by $\mathsf{sunpy}$'s $\mathsf{aiapy.calibrate}$ \citep{Barnes2020} were aligned with an accuracy of less than $1.0 \,\mathrm{arcsec}$.

We analysed small flares with plasma eruptions at chromospheric temperatures captured by SDDI and AIA from 21:56:17 UT on September 6, 2019, to 8:32:40 UT on September 7, 2019.
All of these events occurred in the QS.
We excluded near-limb events and analysed 25 small flares.
These events are the same as our previous study investigating the relationship between chromospheric ejections and small flare energies in the QS \citep{2023ApJ...943..143K}.
As \citet{2023ApJ...943..143K} described, these 25 events were selected for having observed ejecta near the disc centre in the wing difference images obtained by SMART/SDDI.
%Although previous studies have reported that small flares in the QS are often accompanied by dark ejecta \citep{2021ApJ...921L..20P}, note that the results of our analysis are based on the events with dark ejecta.
Previous studies have reported that small flares in the QS are often accompanied by dark ejecta \citep{2021ApJ...921L..20P}.
Our present analysis is specifically focused on events exhibiting dark ejecta, which may limit the generalizability of our findings to other events.
Further investigations are required to establish the broader applicability of our results beyond the specific subset of events analysed in this study.

We used SDO/AIA data to study the properties of small flares.
We checked images of the 304 \AA\, and 6 coronal channels (94 \AA, 131 \AA, 171 \AA, 193 \AA, 211 \AA, and 335 \AA).
The method used to determine the physical quantities (flare spatial scale, temperature, emission measure, and electron number density) of small flares was the same as in our previous study \citep{2023ApJ...943..143K} using differential emission measure (DEM) analysis \citep{2012A&A...539A.146H}.
%We defined the flare duration as the time between the brightening becoming visible and returning to its original intensity in the AIA 193 \AA\, images. 
We defined the flare duration as the time between the brightening beginning and returning to its original intensity in the AIA 193 \AA\, images. 
Thus, even if multiple brightening peaks are included, they are assumed to be the duration of a single event.

We used the \textit{Helioseismic and Magnetic Imager} \citep[HMI: ][]{2012SoPh..275..207S} onboard SDO to study the photospheric magnetic field of the small flares.
We used magnetograms with a 45 seconds cadence because of the short duration of the events.
%We rotated the HMI images to align with the AIA images.

\section{results}

\subsection{Overview}      \label{sec3_1}

\begin{figure*}
	\includegraphics[width=2\columnwidth]{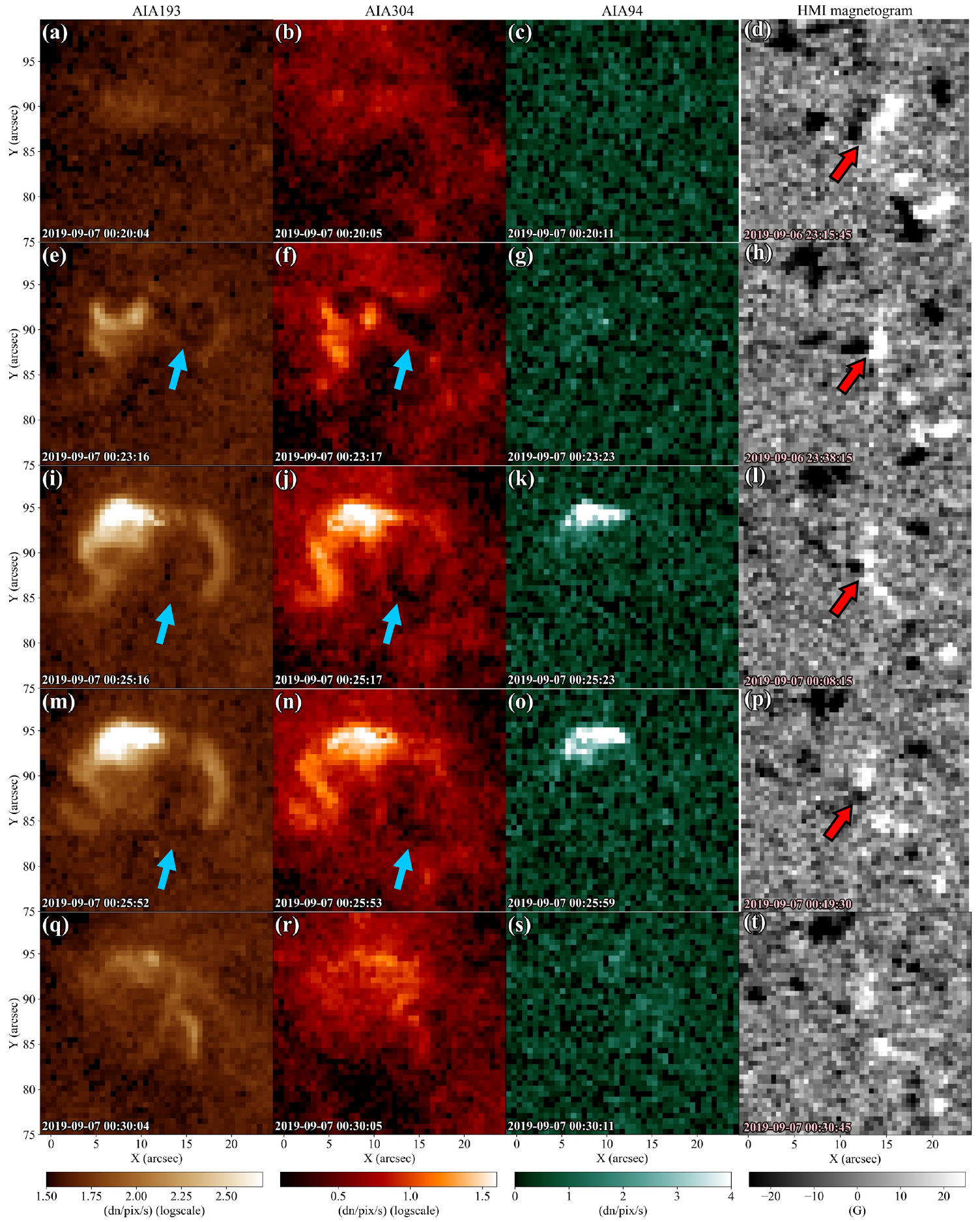}
    \caption{The time evolution of a typical event (event 4 in Table \ref{tab:example_table}) in AIA and HMI images.
    The three left columns show the time evolution in AIA 193 \AA, 304 \AA, and 94 \AA, respectively.
    Light blue arrows indicate dark ejecta.
    The right column shows the time variation of HMI magnetograms.
    Red arrows indicate dipole magnetic fields that indicate cancellation.
    HMI images are corrected for the effect of the solar rotation to track the field of view of the panel (t).
    Note that the AIA and HMI images on the same row are at different times.
    }
    \label{fig:test}
\end{figure*}

In Fig. \ref{fig:test}, we show a typical example of an event analysed in this study.
We can see a brightening appearing in the AIA images from Fig. \ref{fig:test}.
The dark ejecta can be seen in the AIA 193 \AA\, and 304 \AA\, images (Figs. \ref{fig:test}e, f, i, j, m, and n).
The brightening in AIA 94 \AA\, is weaker than in 193 \AA\,(Figs. \ref{fig:test}k and o).
We can see from the HMI images that magnetic field polarities converge and cancel (Figs. \ref{fig:test}d, h, l, and p).
We also confirmed from the broader field of view images that this event occurred at the network boundary.
All these properties are consistent with previous studies \citep[e.g.,][]{2004ApJ...616..578S,2021ApJ...921L..20P}.

\begin{table*}
	\centering
	\caption[Parameters of anlysed events ]{Parameters of small flares analysed in this study.}
	\label{tab:example_table}
	\begin{tabular}{lcccccccccr} % four columns, alignment for each
\hline	
event	 & peak time (UT) & X (arcsec) & Y (arcsec) & $L\,\mathrm{(km)}^{\mathrm{a}}$ & $t_{\mathrm{dur}}\,\mathrm{(s)}^{\mathrm{b}}$ & $\log(T_{\mathrm{DEM}}/[\mathrm{K}])^{\mathrm{c}}$ & $n$ $(10^9\,\mathrm{cm}^{-3})^{\mathrm{d}}$& $v_{\mathrm{red}}$ $(\mathrm{km}\,\mathrm{s}^{-1})^{\mathrm{e}}$ &FC$^{\mathrm{f}}$ & comments \\ \hline
	1 & 9/6 22:02:16  & -350 & 230 & 1100 & 360 &  6.12  &  2.12  &  & Y & MB$^{\mathrm{g}}$ \\
2 & 9/6 23:35:16 & -260 & -360 & 1500 & 360 &  6.14  &  2.54  &  2.82  & Y & jet? \\
3 & 9/7 00:09:16 & -70 & -450 & 2400 & 636 &  6.19  &  3.19  &  4.83  & Y & CBP, jet \\
4 & 9/7 00:25:52 & 10 & 90 & 3800 & 516 &  6.17  &  1.71  &  2.03  & Y &  \\
5 & 9/7 01:32:16 & 320 & 470 & 1000 & 408 &  6.09  &  2.31  &  0.97  & Y & jet, MB \\
6 & 9/7 02:11:52 & 40 & -450 & 2800 & 1008 &  6.17  &  2.94  &  4.48  & Y & CBP \\
7 & 9/7 02:18:16 & 620 & -180 & 2200 & 540 &  6.12  &  1.74  &  & Y &  \\
8 & 9/7 02:25:16 & -420 & 480 & 3600 & 912 &  6.14  &  1.93  &  & Y & MB \\
9 & 9/7 02:32:28 & -540 & -40 & 1200 & 84 &  6.12  &  1.67  &  2.47  & Y &  \\
10 & 9/7 03:17:04 & 610 & 510 & 2700 & 180 &  6.20  &  1.78  &  & Y & CBP, jet, MB \\
11 & 9/7 03:20:28 & -100 & -50 & 3200 & 900 &  6.13  &  1.79  &  & Y & jet, MB \\
12 & 9/7 03:47:28 & -70 & 260 & 4300 & 1044 &  6.17  &  1.55  &  & Y & CBP, jet, MB \\
13 & 9/7 05:28:52 & 200 & -290 & 2700 & 312 &  6.21  &  1.79  &  4.40  & A & CBP \\
14 & 9/7 05:38:52 & -220 & 290 & 2200 & 360 &  6.15  &  1.91  &  1.50  & Y & CBP?, jet, MB \\
15 & 9/7 05:47:28 & 710 & 160 & 6200 & 1188 &  6.15  &  2.96  &  4.22  & Y & CBP, jet, MB \\
16 & 9/7 06:03:16 & 340 & -480 & 1300 & 252 &  6.13  &  2.53  &  5.08  & Y & CBP, jet? \\
17 & 9/7 06:48:28 & -320 & -540 & 1600 & 360 &  6.11  &  2.11  &  & Y &  \\
18 & 9/7 06:50:40 & -420 & 430 & 1100 & 600 &  6.11  &  1.93  &  & Y & MB \\
19 & 9/7 06:49:28 & -30 & -510 & 2800 & 264 &  6.07  &  1.39  &  4.56  & Y & jet, MB \\
20 & 9/7 06:56:52 & 140 & -790 & 2400 & 156  &  6.13  &  1.53  &  & Y & CBP \\
21 & 9/7 07:06:52 & 650 & -100 & 3500 & 548 &  6.13  &  1.46  &  & Y & jet \\
22 & 9/7 07:13:52 & -270 & -660 & 1200 & 408 &  6.11  &  2.80  &  3.50  & A & MB \\
23 & 9/7 07:27:40 & 490 & -30 & 1300 & 216 &  6.16  &  2.44  &  3.95  & Y & MB \\
24 & 9/7 07:35:16 & 190 & 350 & 4600 & 1368 &  6.20  &  1.97  &  4.35  & Y &  \\
25 & 9/7 07:55:52 & -660 & 320 & 1500 & 228 &  6.11  &  2.24  &  & Y & jet \\ \hline
	\end{tabular}
\begin{tablenotes}
\item{(a)} flare spatial scale (square root of the area of brightening pixels)
\item{(b)} flare duration
\item{(c)} temperature in the flare peak time
\item{(d)} electron number density in the flare peak time
\item{(e)} maximum velocity for the redshift associated with H$\alpha$ line centre brightening
\item{(f)} FC = flux cancellation. ``Y" indicates that the cancellation has been confirmed. ``A" indicates ambiguous events.
\item{(g)} MB =multiple brightening.
\end{tablenotes}
\end{table*}

\begin{figure*}
	\includegraphics[width=2\columnwidth]{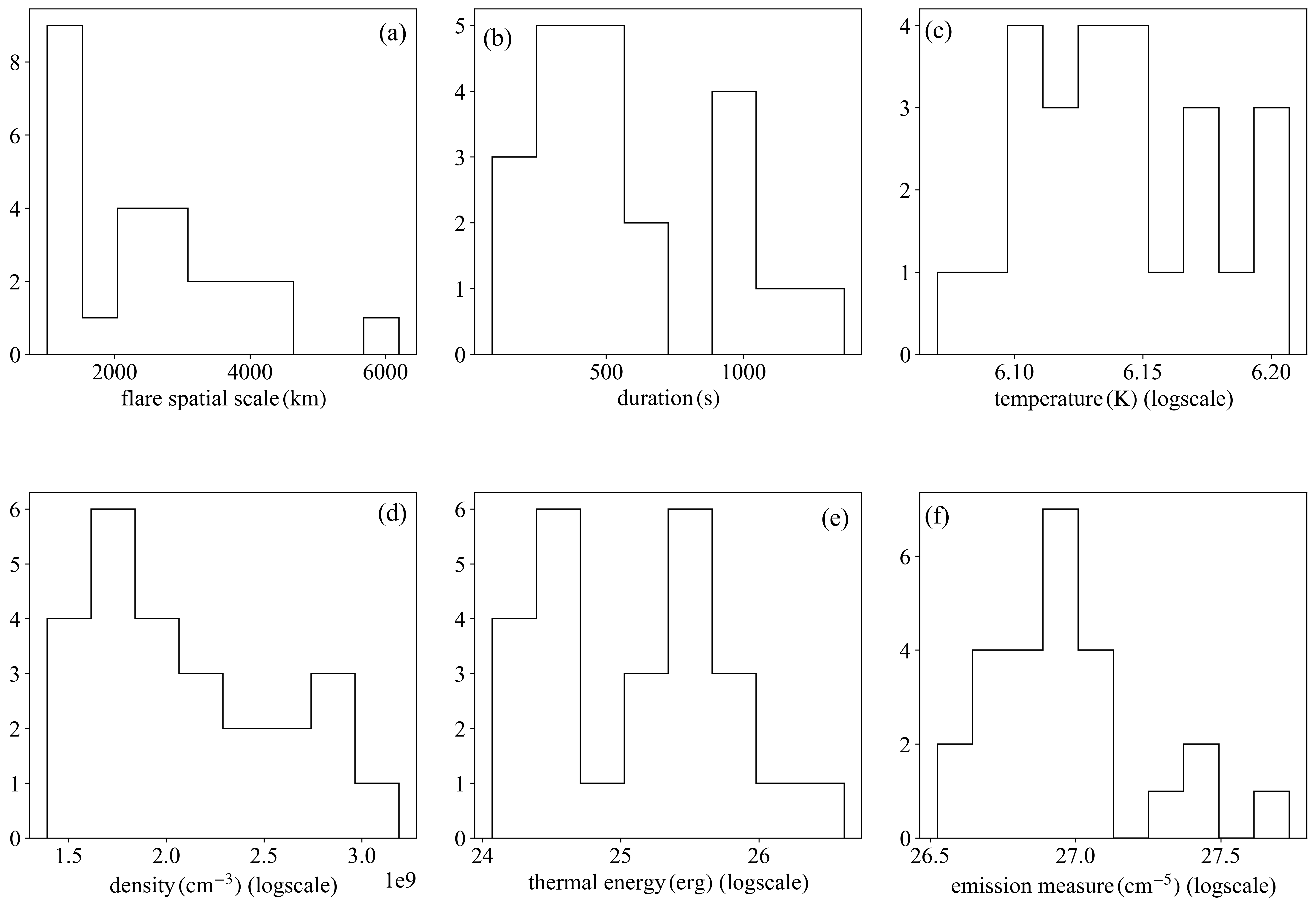}
    \caption{Histogram of the physical quantities of the events analyzed in this study.
    The panels (a), (c), and (f) are the same as in fig. 5 of \citet{2023ApJ...943..143K}.}
    \label{fig:his}
\end{figure*}

We show the physical quantities and their histograms for the events analysed in this study in Table \ref{tab:example_table} and Fig. \ref{fig:his}.
We can see from Fig. \ref{fig:his} that the events have spatial scales $L$ (square root of the area of brightening pixels) between $1000\,\mathrm{km}$ and  $6000\,\mathrm{km}$ and durations $t_{\mathrm{dur}}$ between $100\,\mathrm{s}$ and  $1000\,\mathrm{s}$.
The events have temperatures $T_{\mathrm{DEM}}$ between $10^{6.05}\,\mathrm{K}$ and  $10^{6.2}\,\mathrm{K}$ and emission measure (EM) between $10^{26.5}\,\mathrm{cm}^{-5}$ and  $10^{27.5}\,\mathrm{cm}^{-5}$.
From these values, we can estimate the thermal energies between $10^{24}\,\mathrm{erg}$ and  $10^{26.5}\,\mathrm{erg}$ and electron densities $n$ between $1.5\times10^9\,\mathrm{cm}^{-3}$ and  $3.0\times10^9\,\mathrm{cm}^{-3}$.
Note that kinetic energies of chromospheric temperature ejecta were estimated in our previous study \citep[][Fig. 7]{2023ApJ...943..143K}.
As a result, the kinetic energies tended to be greater than the thermal energies.

Based on these spatial scales and durations, we expect that the events analysed in this study correspond to the smaller events analysed in \citet{2000ApJ...535.1047A}.
Whilst the temperature is consistent with the results of their study, the density is a factor of three to ten times larger in our analysis.
This difference in density is expected due to the different methods used to obtain the DEM.
\citet{2000ApJ...535.1047A} estimated EM from the 195 \AA\, filter of TRACE.
 By contrast, we used AIA 6 channels to estimate DEM distributions for a wider range of temperatures and the sum of these was used as EM. 
 Thus, the EM values estimated in \citet{2000ApJ...535.1047A} are typically ten times smaller than those in our analysis.
The thermal energy values appear to be consistent with \citet{2000ApJ...535.1047A}; however, given the different density values, they may have underestimated the thermal energy by a factor of three to ten.

By comparing the recent SO/EUI campfire observations with the current analysis, we can see that our event corresponds to a large campfire \citep{2021A&A...656L...4B}.
Spatial scale, lifetime, and EM have larger values in our analysis, whilst temperature values were almost the same.
All of our events occurred at the network boundary, and we could clearly see flux cancellation in 92 \% of the events.
These features are consistent with campfires \citep{2021A&A...656L...4B,2021ApJ...921L..20P}.

Some of the events analysed in this study had some distinctive qualitative characteristics.
Nine of the analysed events occurred at CBPs.
It is known that small flares with eruptions of cold plasma also occur in CBPs \citep{2019LRSP...16....2M}.
We should note that the events at CBP tended to have larger temperatures and electron densities but not as large as the microflares in the active region \citep{2008ApJ...677..704H}.
Coronal jets were also observed in several of our events.
Considering that a jet with chromospheric temperature accompanying a coronal jet is formed by magnetic forces  \citep{1996PASJ...48..353Y}, it is a natural result that coronal jets were observed in some of our events, which had chromospheric plasma eruptions.
Some recent campfire observations have also reported the appearance of jets with coronal brightenings \citep{2021A&A...656L..13C}.

\subsection{H$\alpha$ spectrum} \label{sec3_2}

\begin{figure*}
\begin{center}
	\includegraphics[width=2\columnwidth]{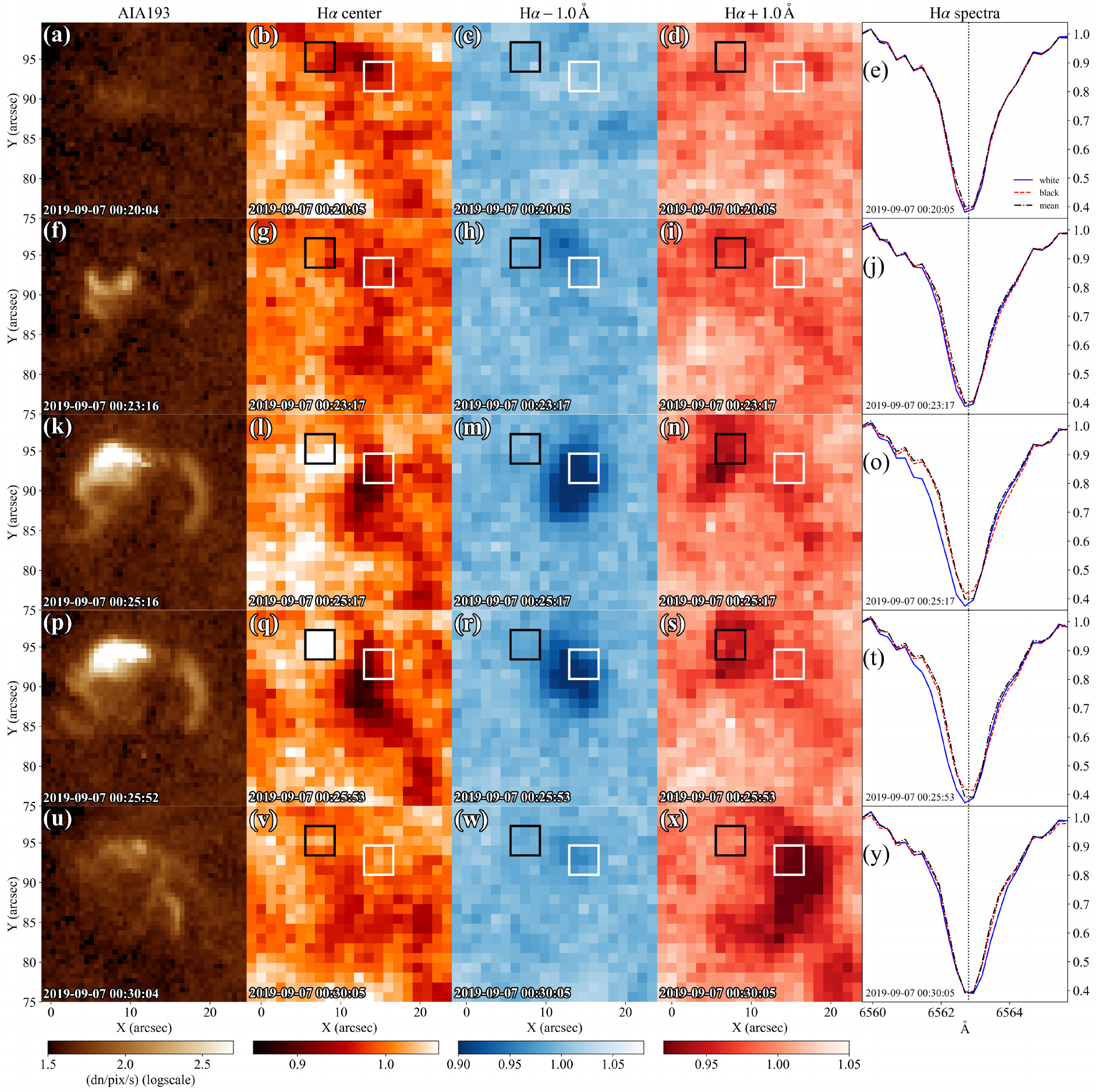}
    \caption{The time evolution of a typical event in the SDDI images and spectra.
    The event is the same as in Fig. \ref{fig:test}.
    The left column shows the AIA 193 \AA\, images shown for comparison.
    The middle columns show the time evolution in the H$\alpha$ line at line centre, $-1.0$ \AA, and $+1.0$ \AA, respectively, from left to right.
    Each image is normalised by the average intensity of the surrounding area.
    Black squares indicate pixels of brightening at line centre with redshift.
White squares indicate pixels of dark ejecta.    
The right column shows the time evolution of the H$\alpha$ spectra.
The solid blue and dashed red lines indicate spectra averaged by white and black squares, respectively.
The dash-dotted black line indicates the average spectra of the surrounding area.
Each spectrum is normalised to the H$\alpha -3.0$ \AA\, intensity of the average spectrum.
    }
    \label{fig:test_sddi}
    \end{center}
\end{figure*}

\begin{figure}
\begin{center}
	\includegraphics[width=\columnwidth]{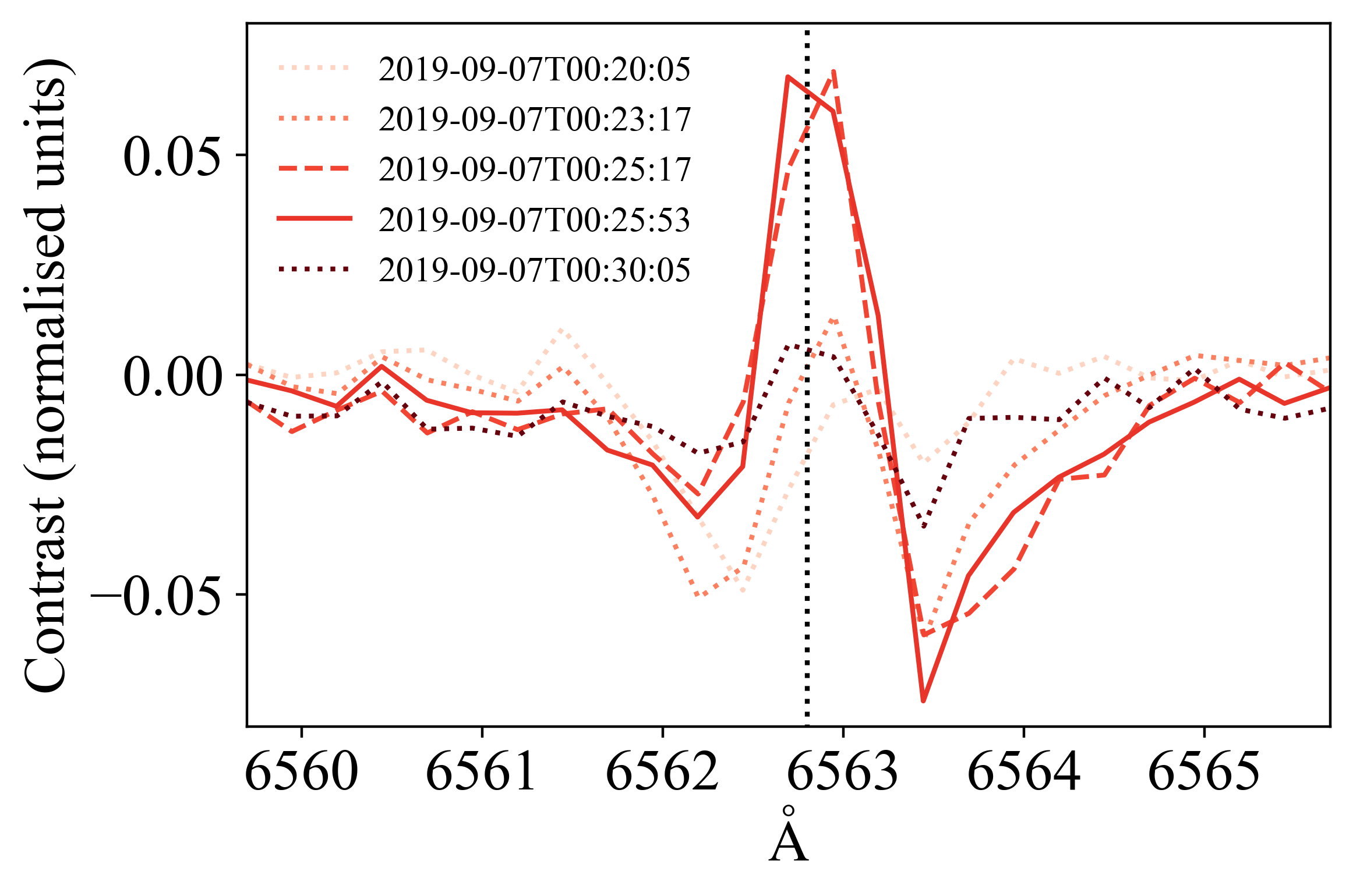}
    \caption{Time evolution of contrast for the redshift associated with the line centre brightening.
    Contrast is obtained as $(I_{\mathrm{red}} - I_0)/I_0$, where $I_{\mathrm{red}}$ is the redshifted spectrum shown as the red line in Fig. \ref{fig:test_sddi} and $I_0$ is the average spectrum shown as the black line.
   Each line shows the contrast at each time of Fig.\ref{fig:test_sddi}.}
    \label{fig:contrast}
    \end{center}
\end{figure}

In Figs. \ref{fig:test_sddi} and \ref{fig:contrast} we show a typical example of a small flare recorded in the AIA 193 \AA\, and H$\alpha$ line.
We can see a brightening in the corresponding pixels in the H$\alpha$ line centre at the time when the brightness in AIA 193 \AA\, reaches its peak (Figs. \ref{fig:test_sddi}l and q black squares).
This enhancement is insignificant as the intensity increases by only about 5 \% over the ambient intensity.
This brightening is seen in the H$\alpha$ images taken at $+1.0$ \AA\, as a dark structure; that is, it is associated with H$\alpha$ redshift (Figs. \ref{fig:test_sddi}n, o, s, and t).
We show its characteristics more explicitly by calculating the contrast of the redshifted spectra (Fig. \ref{fig:contrast}).
In addition, the dark ejecta identified in the AIA images is also visible as absorption in the H$\alpha$ line centre (Figs. \ref{fig:test_sddi}l and q white sqares).
This ejecta is also seen in the H$\alpha -1.0$ \AA; thus, it shows blueshifted absorption(Figs. \ref{fig:test_sddi}m, o, r, and t).
These results confirm that the dark ejecta identified in the AIA images contain plasma at chromospheric temperatures.
After the brightening and blueshift disappear, another redshifted absorption appears near the pixel where the ejecta was originally in the H$\alpha +1.0$ \AA\, (Figs. \ref{fig:test_sddi}x and y).
Based on the spatial and temporal consistency with the ejecta observed in the blueshift, this redshift would be the ejecta falling due to gravity.

We found a redshift associated with line centre brightening in the H$\alpha$ line corresponding to AIA brightening in 14 cases.
Previous studies often reported the enhancement at the line centre associated with minifilament eruption  \citep{1986NASCP2442..369H,2004ApJ...616..578S,2020A&A...643A..19M}.
However, the redshift with the line centre brightening has rarely been observed in chromospheric spectra \citep{2020ApJ...898..144K,2022A&A...660A..45M,2022ApJ...939...25P}, and this is the first time it has been investigated in many events.
This result indicates that even small flares in the QS can affect the chromosphere.
The reason that the brightening and redshift were observed in only half of the events could be attributed to the insufficient spatial resolution of the SDDI. 
Another reason could be that the energy flux injected into the chromosphere is not extremely large compared to the radiation flux in the chromosphere.
We can estimate the thermal conduction flux $F_{\mathrm{cond}}$ (in units of $\,\mathrm{erg}\,\mathrm{cm}^{-2}\,\mathrm{s}^{-1}$) using the typical physical quantities of the events analysed in this study as follows:
\begin{equation}
F_{\mathrm{cond}}\sim \kappa_0 \frac{T_{\mathrm{DEM}}^{7/2}}{L} \simeq 9.0\times 10^6\left(\frac{T_{\mathrm{DEM}}}{10^{6.1}\,\mathrm{K}} \right)^{7/2} \left(\frac{L}{2.5\times10^{8}\,\mathrm{cm}} \right)^{-1}       , \label{F_thcond_obs}
\end{equation}
where $\kappa_0\simeq 10^{-6}\,\mathrm{cgs}$ is the Spitzer thermal conductivity.
This value is larger than the energy flux in the chromosphere \citep[upper chromosphere: $3\times10^5\,\mathrm{erg}\,\mathrm{cm}^{-2}\,\mathrm{s}^{-1}$, middle chromosphere: $2\times10^6\,\mathrm{erg}\,\mathrm{cm}^{-2}\,\mathrm{s}^{-1}$,][]{1977ARA&A..15..363W} but much smaller than the energy flux of a typical solar flare ($>10^{9}\,\mathrm{erg}\,\mathrm{cm}^{-2}\,\mathrm{s}^{-1}$).

\begin{figure}
\begin{center}
	\includegraphics[width=\columnwidth]{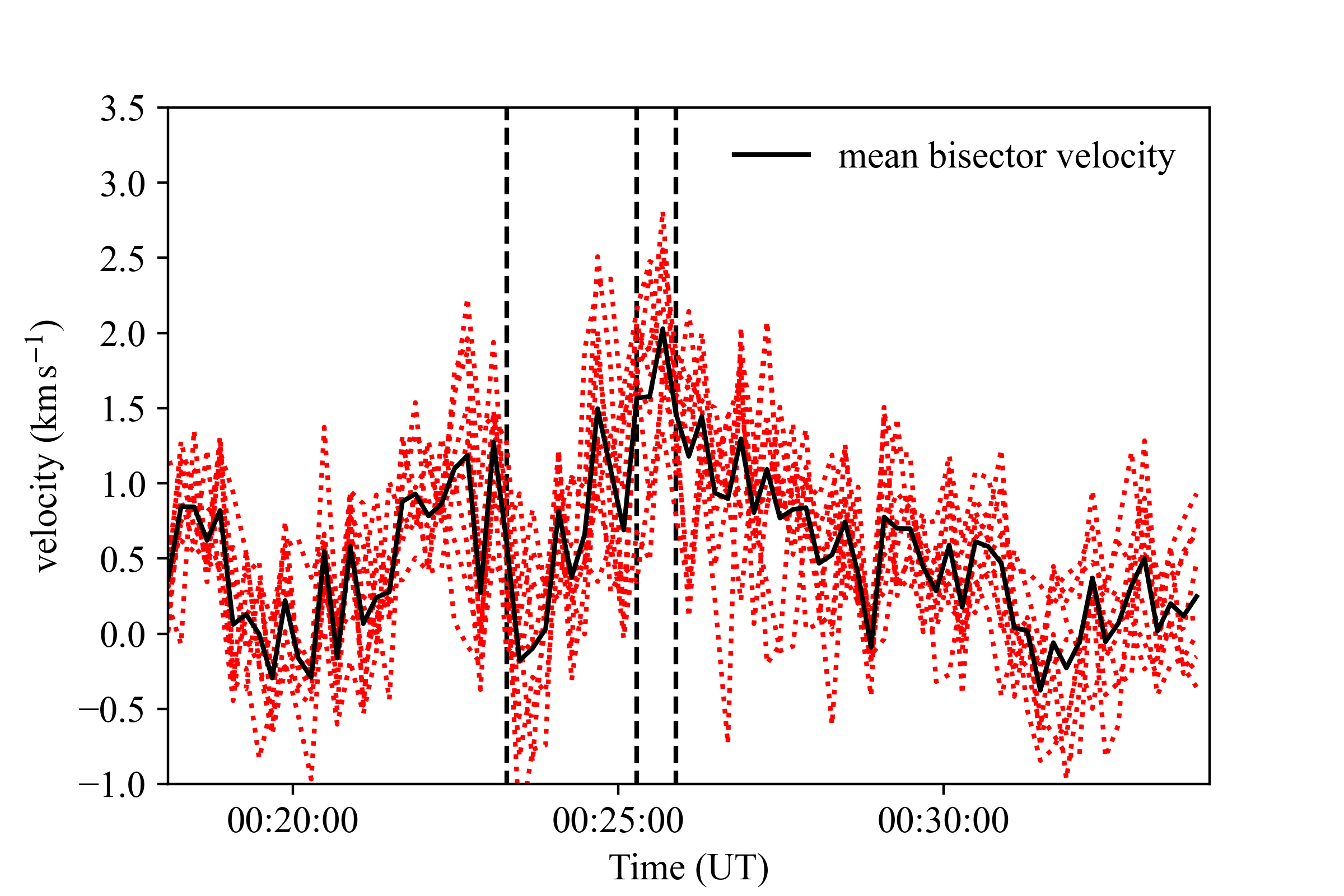}
    \caption{Typical example of redshift velocity with brightening in H$\alpha$ line centre.
        The event is the same as in Fig. \ref{fig:test}.
        The dashed black vertical lines indicate the times in Figs. \ref{fig:test}e, i, and m, respectively.
        The dotted red lines indicate the time evolution at each of the pixels surrounded by the black squares in Fig. \ref{fig:test_sddi}.
        The solid black line shows the average of the dotted red lines.
    }
    \label{fig:test_bis}
    \end{center}
\end{figure}

We determined the bisector velocity to characterise the redshift associated with the brightening of the H$\alpha$ line centre \citep{1966ApJ...146..194K}.
We obtained the intensity $I_{\mathrm{bis}}$ for determining the bisector velocity as follows:
\begin{equation}
I_{\mathrm{bis}} =  I_{\mathrm{H\alpha}}(0\,\text{\AA}) + 0.35\left( \frac{I_{\mathrm{H\alpha}}(+3.0\,\text{\AA}) + I_{\mathrm{H\alpha}}(-3.0\,\text{\AA})}{2} - I_{\mathrm{H\alpha}}(0\,\text{\AA})\right)    , \label{I_bis}
\end{equation}
where $ I_{\mathrm{H\alpha}}(\lambda\,\text{\AA})$ is the H$\alpha$ intensity at wavelengths shifted by $\lambda$ \AA\, from the line centre.
%\textcolor{blue}{With the difference between the two wavelengths corresponding to the intensity $I_{\mathrm{bis}}$ as $\Delta\lambda$, we determined the bisector velocity $v_{\mathrm{red}}$ as follows:
With the two wavelengths corresponding to the intensity $I_{\mathrm{bis}}$ as $\lambda_+$ and $\lambda_-$, we determined the bisector velocity $v_{\mathrm{red}}$ as follows:
\begin{equation}
%v_{\mathrm{red}}=  \frac{\Delta\lambda}{\lambda_0}c , \label{bis_vel_def}
v_{\mathrm{red}}=  \frac{0.5(\lambda_+ + \lambda_-) - \lambda_0}{\lambda_0}c , \label{bis_vel_def}
\end{equation}
where $\lambda_0= 6562.8$ \AA\, is the H$\alpha$ line centre wavelength and $c=3.0\times10^5\,\mathrm{km}\,\mathrm{s^{-1}}$ is the light speed.
The bisector velocity corresponding to the intensity $I_{\mathrm{bis}}$ was determined for each of the 9 pixels where brightening occurred, as shown in Fig. \ref{fig:test_sddi}.
We then took the average of these 9 pixels.
We show the typical behaviour of the bisector velocity obtained at each time in Fig. \ref{fig:test_bis}.
We can see from Fig. \ref{fig:test_bis} that the bisector velocity increases with the brightening.
We focused on the average bisector velocity of 9 pixels, and the maximum value of its time evolution was taken as the bisector velocity of the event.

We summarise the bisector velocity $v_{\mathrm{red}}$ for each event in Table \ref{tab:example_table}.
The events have bisector velocities between $1.0\,\mathrm{km}\,\mathrm{s}^{-1}$ and $5.0\,\mathrm{km}\,\mathrm{s}^{-1}$.
These values are slightly larger than the redshift velocity in the steady chromosphere \citep{1998ApJS..114..151C}.

\subsection{Light Curve}      \label{sec3_3}

\begin{figure}
\begin{center}
	\includegraphics[width=\columnwidth]{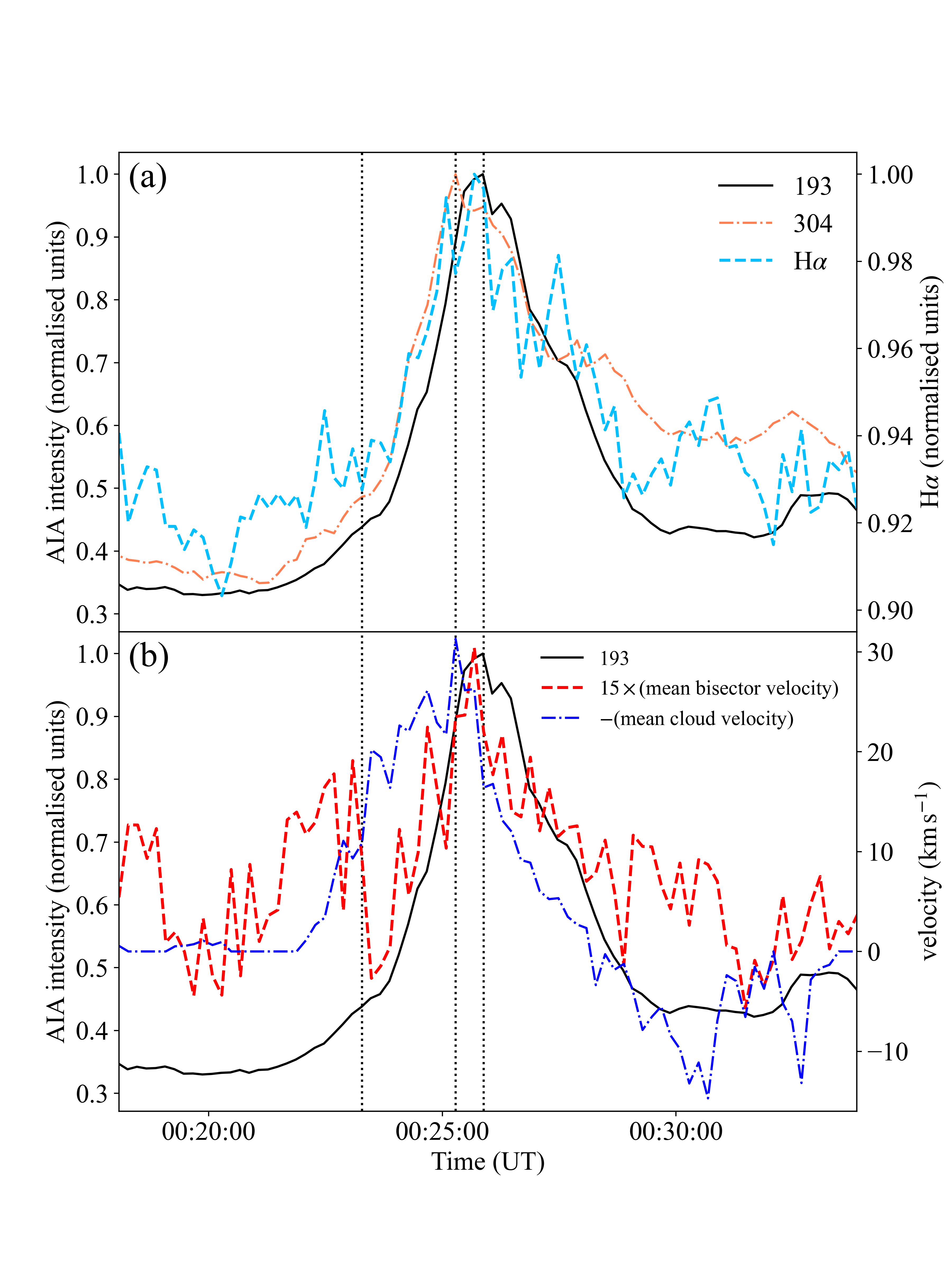}
    \caption{Typical examples of the light curves.
     The event is the same as in Fig. \ref{fig:test}.
     The dotted black vertical lines indicate the times in Figs. \ref{fig:test}e, i, and m, respectively.
     (a): AIA 193 \AA\,and 304 \AA, and H$\alpha$ line centre light curves.
     The solid black, dashed light blue, and dash-dotted orange lines indicate the light curves of 193 \AA, 304 \AA, and H$\alpha$ line centre, respectively.
     (b): Comparison of AIA 193 \AA\, light curves with redshift velocities with brightening in H$\alpha$ centre, and velocities of the dark ejecta.
     The dashed red and dash-dotted blue lines indicate the velocity of the redshift and the dark ejecta, respectively.
     The redshift velocity is shown multiplied by a factor of 15.
     The velocity of the dark ejecta is shown with the sign reversed.
     Note that the direction toward the observer is defined as a negative velocity.
    }
    \label{fig:light_curve}
    \end{center}
\end{figure}

In Fig. \ref{fig:light_curve}a we show a typical example of a light curve for the analysed events in this study.
We obtain the AIA light curves as the sum of the pixels showing brightenings. 
The light curves are normalised to the maximum value.
For the H$\alpha$ light curve, we use the sum of the intensities of the nine pixels shown in Fig. \ref{fig:test_sddi}.
The H$\alpha$ light curve is also normalised by its maximum value.
We can see from Fig. \ref{fig:light_curve}a that H$\alpha$ and AIA 304 \AA\, show an earlier intensity increase than AIA 193 \AA.
This property is consistent with previous studies \citep{1999A&A...341..286B,2021ApJ...914L..35J,2022A&A...660A..45M} and is reminiscent of the Neupert effect \citep{1968ApJ...153L..59N}.
Fig. \ref{fig:light_curve}a also shows that the brightening in the H$\alpha$ line varies only about 10 \% over time.

Fig. \ref{fig:light_curve}b shows the time evolution of the redshifted velocity associated with line centre brightening $v_{\mathrm{red}}$ and the velocity of the chromospheric ejecta compared to the AIA 193 \AA\, light curve.
We used the cloud model \citep{1964PhDT........83B,1988A&A...203..162M} and obtained the line-of-sight velocities of the chromospheric ejecta. 
The cloud model is based on modelling the spectrum of a plasma blob above the background radiation field.
That model estimates line-of-sight velocity $v_l$, the optical thickness of the blob $\tau_0$, source function $S$, and Doppler width $\Delta\lambda_D$ by fitting the contrast $C(\lambda)$ using the following equation:
%\begin{equation}
%C(\lambda) = \frac{I_{\mathrm{H\alpha}}(\lambda) - I_{\mathrm{H\alpha,0}}(\lambda)}{I_{\mathrm{H\alpha,0}}(\lambda)} = \left( \frac{S}{I_{\mathrm{H\alpha,0}}(\lambda)} - 1\right)\left( 1- e^{-\tau(\lambda)}\right), \label{cloud_model}
%\end{equation}
%}
\begin{equation}
C(\lambda) = \frac{I(\lambda) - I_0(\lambda)}{I_0(\lambda)} = \left( \frac{S}{I_0(\lambda)} - 1\right)\left( 1- e^{-\tau(\lambda)}\right), \label{cloud_model}
\end{equation}
\begin{equation}
\tau(\lambda) = \tau_0 \exp\left( -\frac{\lambda -\frac{v_l}{c}\lambda_0}{\Delta\lambda_D} \right), \label{cloud_tau}
\end{equation}
where $I(\lambda)$ is spectra of the plasma blob, $ I_0(\lambda)$ is the background radiation, and $\lambda_0$ is a wavelength at the line centre.
We used the average spectrum of the surroundings as $ I_0(\lambda)$.
We averaged the line-of-sight velocity over the nine pixels shown in Fig. \ref{fig:test_sddi} and examined its temporal evolution. 
Fig. \ref{fig:light_curve}b shows that both the ejecta and the redshifted velocity peaked before the 193 \AA\, light curve peak.
The ejecta accelerates before the small flare reaches its peak, reaching a maximum value at the same time as the 304 \AA\, light curve.
This property is consistent with the standard flare model in which mass ejection triggers magnetic reconnection \citep{1998ApJ...499..934O,1995ApJ...451L..83S}.
The redshift velocity is highly variable, and we cannot clearly identify an acceleration phase.
However, its peak is synchronised with the H$\alpha$ line light curve.
This fact provides stronger support that the redshift is accompanied by the line centre brightening.

%繰り返し増光が起きていたものの話

\section{discussion}

\subsection{Comparison of H$\alpha$ redshifted spectra and chromospheric condensation theory} \label{sec4_1}

Section \ref{sec3_2} confirmed that H$\alpha$ line spectra show the line-centre brightening and redshift synchronised with the coronal brightening.
Here we discuss the possibility that this redshift corresponds to chromospheric condensation in the small flares in the QS.

The line centre brightening and redshift of the H$\alpha$ line found in our study are reminiscent of the red asymmetry observed in large-scale flares \citep{1984SoPh...93..105I}.
In large flare cases, released flare energy is injected into the chromosphere from the corona, resulting in chromospheric condensation \citep{1985ApJ...289..434F}.
The propagation of the condensation shock downward is observed as red asymmetry.
The spectra found in this study are not emission lines; however, they are similar to the red asymmetry of large flares in that they show a redshift associated with line centre brightening.
This difference is probably because the source function of the H$\alpha$ line increases due to energy injection from the corona even for small flares in the QS, but the resulting brightening is not large enough for the H$\alpha$ line to be observed in emission.

Some synthesised H$\alpha$ profiles corresponding to chromospheric condensation originating from a nanoflare have been reported to have a similar profile to the present observation.
\citet{2022A&A...659A.186B} performed numerical simulations for nanoflares in active regions and produced synthesised profiles of the H$\alpha$ line.
As a result, they reproduced the H$\alpha$ line spectra, which remain absorption lines but show an increase in the line centre intensity and a redshift \citep[see][fig. 7]{2022A&A...659A.186B}.
However, their synthesised H$\alpha$ spectra also show a redshift at the line centre.
This is different from our observations, which show almost no shift in the line centres and a redshift only around $+1.0$ \AA.
Their calculations also report a phase in which the H$\alpha$ line becomes an emission line before becoming an absorption line, but we found no H$\alpha$ emission line in our observation.
One reason for these differences is that the spatial and temporal resolution of SDDI (pixel size: 1.23 arcsec/temporal  resolution: 12 s) is not as good as the numerical simulations.
In other words, we should observe a superposition of spectra from multiple loops, resulting in a difference from the synthetic observation calculated for a single loop.
Another reason may be that \citet{2022A&A...659A.186B} calculated using different parameters than the small loops in the QS.
\citet{2022A&A...659A.186B} synthesised the H$\alpha$ line with 1D radiative transfer and did not incorporate the 3D effects, which are essential for the formation of H$\alpha$ lines \citep{2012ApJ...749..136L}.
This may be one of the reasons why the H$\alpha$ lines do not completely agree with observations.

\begin{figure}
	\includegraphics[width=\columnwidth]{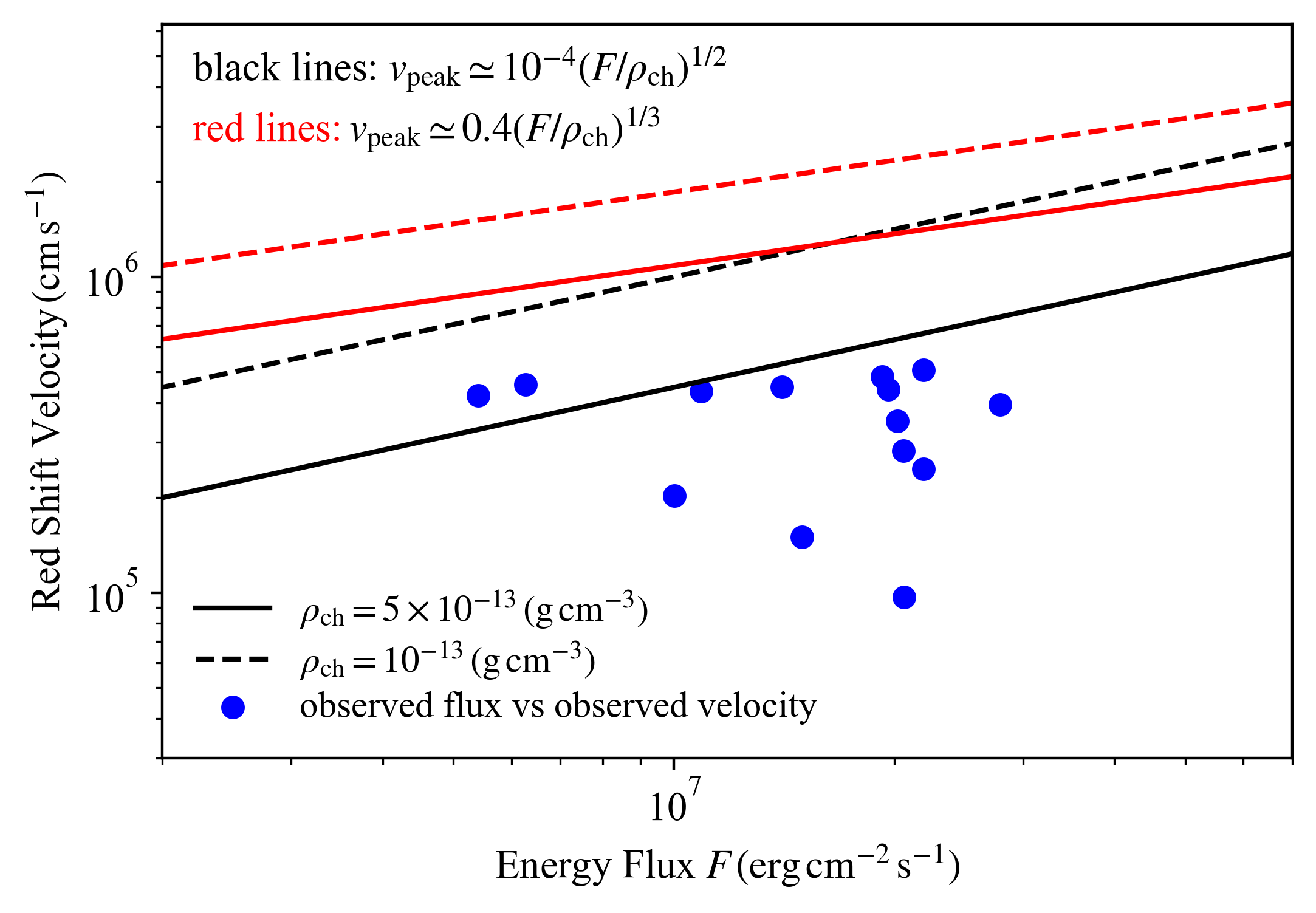}
%	\epsscale{0.75}
%	\plotone{EM_T_9_11.png}
    \caption{Comparison of the observed redshift velocity associated with brightening in the H$\alpha$ centre with the formula for chromospheric condensation \citep{1989ApJ...346.1019F,2014ApJ...795...10L}.
    The black and red lines indicate the \citet{2014ApJ...795...10L} formula (equation \ref{Long2014}) and the \citet{1989ApJ...346.1019F} formula (equation \ref{Fisher1989}), respectively.
    The solid and dashed lines show the equations (\ref{Fisher1989}) and (\ref{Long2014}) assuming different mass densities of the chromosphere, respectively.
    }
    \label{fig:vred_F}
\end{figure}

%not including Fisher version

%We compared the observation result with the theoretical formula to verify that the observed spectra are chromospheric condensation.
%Several relationships have been proposed between the maximum velocity of chromospheric condensation and the energy flux \citep[e.g.,][]{1989ApJ...346.1019F}.
%\citet{2021ApJ...912...25A} performed 1D hydrodynamic  simulations to study the theoretical properties of chromospheric condensation.
%They reported that the formula obtained in \citet{2014ApJ...795...10L} below is consistent for low-energy cases. 
%\begin{equation}
%v_{\mathrm{peak}}\simeq 10^{-4} \left(\frac{F}{\rho_{\mathrm{ch}}} \right)^{1/2}, \label{Long2014}
%\end{equation}
%where $v_{\mathrm{peak}}$ is the condensation peak velocity (in units of $\mathrm{cm}\,\mathrm{s}^{-1}$), $F$ is the injected energy flux into the chromosphere (in units of $\mathrm{erg}\,\mathrm{cm^{-2}}\,\mathrm{s}^{-1}$), and $\rho_{\mathrm{ch}}$ is the mass density in the chromosphere (in units of $\mathrm{g}\,\mathrm{cm^{-3}}$).
%Hence, we compared equation (\ref{Long2014}) with the observations.

%including Fisher version

We compared the observational result with the theoretical formula to verify that the observed spectra are chromospheric condensation.
Following two relationships have been proposed between the maximum velocity of chromospheric condensation and the energy flux \citep[][in order]{1989ApJ...346.1019F,2014ApJ...795...10L}:
\begin{equation}
v_{\mathrm{peak}}\simeq 0.4 \left(\frac{F}{\rho_{\mathrm{ch}}} \right)^{1/3}, \label{Fisher1989}
\end{equation}
\begin{equation}
v_{\mathrm{peak}}\simeq 10^{-4} \left(\frac{F}{\rho_{\mathrm{ch}}} \right)^{1/2}, \label{Long2014}
\end{equation}
%where $v_{\mathrm{peak}}$ is the condensation peak velocity (in units of $\mathrm{cm}\,\mathrm{s}^{-1}$), $F$ is the injected energy flux into the chromosphere (in units of $\mathrm{erg}\,\mathrm{cm^{-2}}\,\mathrm{s}^{-1}$), and $\rho_{\mathrm{ch}}$ is the \textcolor{red}{pre-flare mass density in the upper chromosphere} (in units of $\mathrm{g}\,\mathrm{cm^{-3}}$).
where $v_{\mathrm{peak}}$ is the condensation peak velocity (in units of $\mathrm{cm}\,\mathrm{s}^{-1}$)and $\rho_{\mathrm{ch}}$ is the pre-flare mass density in the upper chromosphere (in units of $\mathrm{g}\,\mathrm{cm^{-3}}$).
$F$ is the injected energy flux into the condensation shock (in units of $\mathrm{erg}\,\mathrm{cm^{-2}}\,\mathrm{s}^{-1}$) in equation (\ref{Fisher1989}) and flare energy flux in equation (\ref{Long2014}).
\citet{1989ApJ...346.1019F} derived equation (\ref{Fisher1989}) in an analytical approach, whilst \citet{2014ApJ...795...10L} derived equation (\ref{Long2014}) empirically from the results of hydrodynamic simulations for a flare loop.
\citet{2014ApJ...795...10L} said that the reason for the different powers in equations (\ref{Fisher1989}) and (\ref{Long2014}) is due to systematic variations in the fraction of the total energy flux reaching condensation shock.

%Hence, we compared equations  (\ref{Fisher1989}) and (\ref{Long2014}) with the observations.

%\citet{2021ApJ...912...25A} performed 1D hydrodynamic  simulations to study the theoretical properties of chromospheric condensation.}
%They reported that the formula obtained in \citet{} below is consistent for low-energy cases. 

Fig. \ref{fig:vred_F} compares equations (\ref{Fisher1989}) and (\ref{Long2014}) and our observations.
Here, we assume the thermal conduction fluxes (equation \ref{F_thcond_obs}) for the energy flux $F$.
We can see from Fig. \ref{fig:vred_F} that equation (\ref{Long2014}) agrees with the upper bound of the observed velocities.
Considering that equation (\ref{Long2014}) corresponds to the maximum condensation velocity, this result supports the interpretation that the spectral variations are due to the chromospheric condensation.
In addition, the better agreement of the \citet{2014ApJ...795...10L} formula than the \citet{1989ApJ...346.1019F} formula is consistent with the case of small-energy flares in the numerical simulation of \citet{2021ApJ...912...25A}.

Another candidate for heated downflow is reconnection outflow, that is, plasma flow accelerated in a current sheet.
However, it is unlikely that the observed spectral variations are due to reconnection outflows for the following two reasons.
First, the observed redshift velocity is small compared to the Alfv\'{e}n velocity in the chromosphere.
Assuming typical parameters for the upper chromosphere, we can estimate the Alfv\'{e}n velocity ($v_A$, in units of $\mathrm{cm}\,\mathrm{s^{-1}}$) as follows:
\begin{equation}
v_{A}=\frac{B_{\mathrm{ch}}}{\sqrt{4\pi\rho_{\mathrm{ch}}}} = 4.0\times10^6\left(\frac{B_\mathrm{ch}}{10\,\mathrm{G}} \right)\left(\frac{\rho_\mathrm{ch}}{5\times10^{-13}\,\mathrm{g}\,\mathrm{cm^{-3}}} \right), \label{alfven_vel}
\end{equation}
where $B_\mathrm{ch}$ is the magnetic field strength in the upper chromosphere (in units of $\mathrm{G}$).
This value is more than ten times larger than our observation.
Second, condensation downflow should be easier to observe than reconnection outflow because of its larger spatial scale.
Reconnection outflow is collimated thin and stops when it collides with the lower loop.
In contrast, the condensation downflow propagates along the top of the lower loop.
Hence, it should have a larger spatial scale than the reconnection outflow in a typical magnetic field morphology.
Based on the above discussion, although reconnection outflow may be mixed, it is likely that condensation downflow mainly contributed to the H$\alpha$ spectral variability.

\subsection{Emission measure vs temperature Scaling Law}     \label{sec4_2}

\begin{figure}
	\includegraphics[width=\columnwidth]{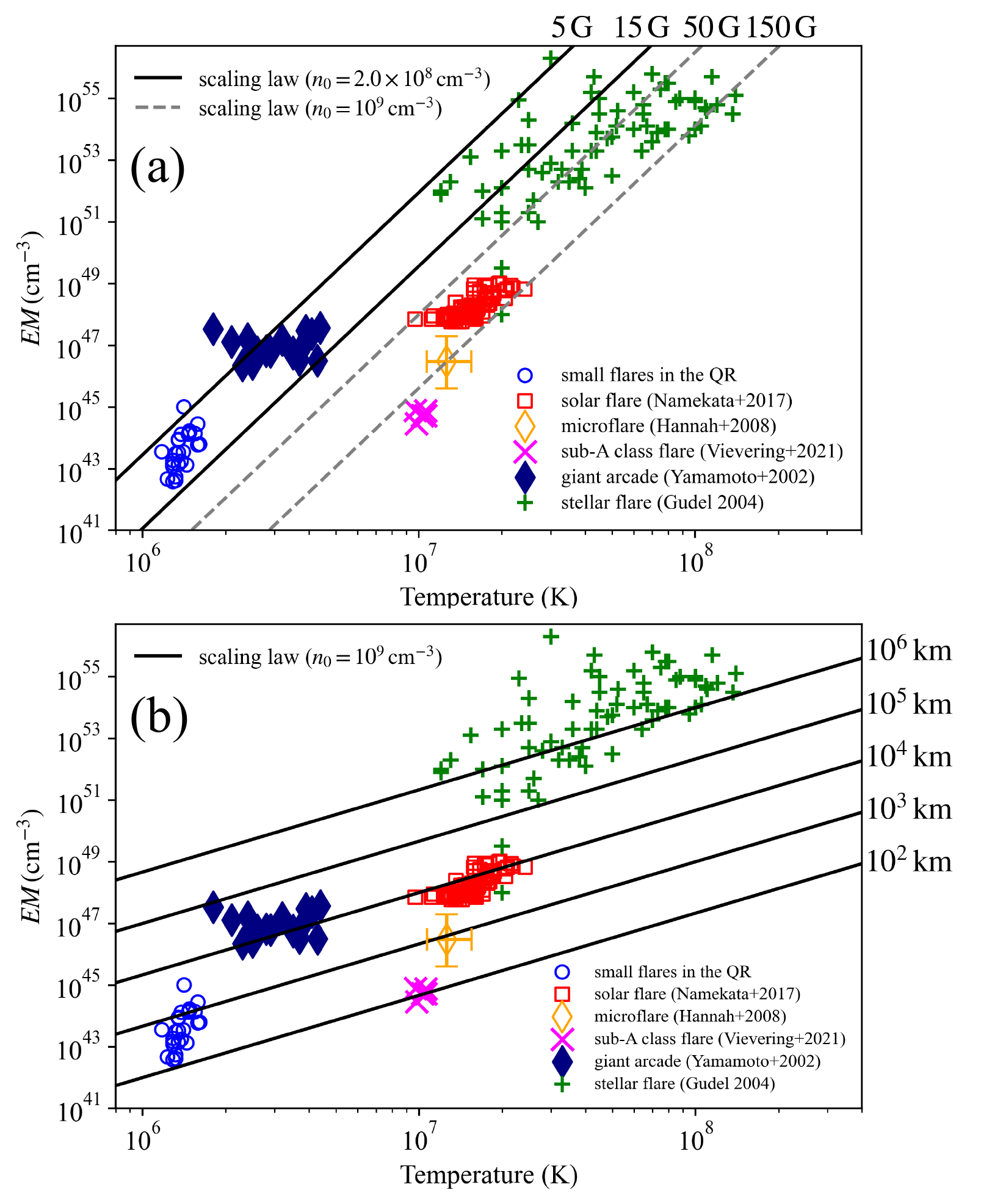}
    \caption{Comparison with EM vs temperature scaling law at time of flare peak \citep{1999ApJ...526L..49S,2002ApJ...577..422S}.
    The blue circles indicate the analysed results of the small flares in the QS in this study.
    Red squares indicate typical solar flares \citep{2017PASJ...69....7N}.
    The orange diamonds and the pink x-marks indicate microflares \citep{2008ApJ...677..704H} and sub-A class flares \citep{2021ApJ...913...15V} in active regions, respectively.
    The dark blue diamonds indicate large flares in the QS \citep[giant arcades,][]{2002ApJ...579L..45Y}.
    Green plus signs indicate stellar flares \citep{2004A&ARv..12...71G}.
    (a): Comparison with the scaling law (equation \ref{SY_scaling}) with the coronal magnetic field strength as a parameter.
    The solid black and dashed gray lines assume $n_0=2\times10^8\,\mathrm{cm}^{-3}$ and $n_0=10^9\,\mathrm{cm}^{-3}$ for the pre-flare electron density in the corona $n_0$, respectively.
    (b): Comparison with the scaling law (equation \ref{SY_scaling_L}) with the spatial scale as a parameter.
        Since in equation (\ref{SY_scaling_L}) $n_0$ only affects the EM by a power-law index of $2/3$, we assume $n_0=10^9\,\mathrm{cm}^{-3}$ for all lines in this figure.
   }
    \label{fig:EM_T}
\end{figure}

To understand the physical mechanism of small flares in the QS, we compared the present DEM analysis results with \citet{1999ApJ...526L..49S,2002ApJ...577..422S} scaling law. 
\citet{1999ApJ...526L..49S,2002ApJ...577..422S} derived the following scaling law between the volume emission measure $EM_v$ ($\mathrm{cm}^{-3}$) and temperature in the flare peak times:
\begin{equation}
EM_v \simeq 10^{48} \left( \frac{B_{\mathrm{corona}}}{50\,\mathrm{G}} \right)^{-5} \left( \frac{n_0}{10^9\,\mathrm{cm}^{-3}} \right)^{3/2}\left( \frac{T_{\mathrm{peak}}}{10^7\,\mathrm{K}} \right)^{17/2}, \label{SY_scaling}
\end{equation}
where $B_{\mathrm{corona}}$ is the coronal magnetic field strength, $n_0$ is the preflare electron number density, and $T_{\mathrm{peak}}$ is the temperature in the flare peak time.
This scaling law is consistent with a wide range of energies, from microflares in active regions to giant stellar flares.

Fig. \ref{fig:EM_T} shows the comparison of our analysis result with the scaling law of \citet{1999ApJ...526L..49S,2002ApJ...577..422S}.
Here, the volume emission measure $EM_v$ is calculated from the observations as follows:
\begin{equation}
EM_v = EM\times L^2 \label{EMv_obs}
\end{equation}
We can see from Fig. \ref{fig:EM_T}a that the present observation agrees with the case where $B_{\mathrm{corona}}=5-15\,\mathrm{G}$ and $n_0=2\times10^8\,\mathrm{cm}^{-3}$ in equation (\ref{SY_scaling}).
We also include in Fig. \ref{fig:EM_T} observations of giant arcades \citep{2002ApJ...579L..45Y} for further understanding.
The giant arcades are also located away from the group of active region flares and agree with equation  (\ref{SY_scaling}) for $B_{\mathrm{corona}}=5-15\,\mathrm{G}$ and $n_0=2\times10^8\,\mathrm{cm}^{-3}$.
These results indicate that \citet{1999ApJ...526L..49S,2002ApJ...577..422S} scaling law can explain small flares in the QS and giant arcades by considering appropriate values of the magnetic field strength and coronal electron density in the QS.

\citet{1999ApJ...526L..49S,2002ApJ...577..422S} also derived the scaling law using the spatial scale $L$ rather than the magnetic field $B_{\mathrm{corona}}$ as a parameter.
\begin{equation}
EM_v \simeq 10^{48} \left( \frac{L}{10^9\,\mathrm{cm}} \right)^{5/3} \left( \frac{n_0}{10^9\,\mathrm{cm}^{-3}} \right)^{2/3}\left( \frac{T_{\mathrm{peak}}}{10^7\,\mathrm{K}} \right)^{8/3}. \label{SY_scaling_L}
\end{equation}
We show in Fig. \ref{fig:EM_T}b our analysis results and equation (\ref{SY_scaling_L}).
We can see a good agreement with the case where $L\simeq10^{3}\,\mathrm{km}$.
We can also see that some of the small flares in the QS are consistent with equation (\ref{SY_scaling_L}) for the case smaller than $1000\,\mathrm{km}$.
These spatial scales are less than the thickness of the chromosphere, which may seem to be a contradictory result.
However, we suggest that these spatial scales reflect that only a portion of the loops brightens at coronal temperatures.
This hypothesis is consistent with recent "campfire'' studies \citep{2021A&A...656L...4B,2021A&A...656A..35Z}. %ここはfilling factorが足りていない可能性も考慮してもう少し深く議論したい、、、

\citet{2008ApJ...672..659A} also compared the $EM_v$ and temperature of small flares in the QS with active region flares.
They obtained the following relationship by a linear regression fit for various scale solar flares:
\begin{equation}
EM_v \simeq 10^{48.4} \left( \frac{T_{\mathrm{peak}}}{10^7\,\mathrm{K}} \right)^{4.7\pm0.1}. \label{Asch_scaling}
\end{equation}
The power-law index in this equation differs from those in equations (\ref{SY_scaling}) and (\ref{SY_scaling_L}).
The following two reasons can explain this difference.
First, \citet{2008ApJ...672..659A} performed the fit without including sub-A class flares in active regions and giant arcades.
Equation (\ref{Asch_scaling}) is clearly inconsistent with these events shown in Fig. \ref{fig:EM_T}.
The second reason is that the fitting in \citet{2008ApJ...672..659A} did not consider differences in the physical quantities associated with flares.
They performed the fitting simultaneously for flares in quiet and active regions where the magnetic field strength differs.
The difference in the magnetic field strength affects the Poynting flux that heats the atmosphere associated with the flare, causing differences in the temperature for the same EM.
Hence, the power index should be smaller than the value in equation (\ref{SY_scaling}) ($=17/2$).
From the above discussion, we can say that equation  (\ref{Asch_scaling}) is the result of fitting by focusing on only some of the many flares that occur in the solar atmosphere without considering the differences in their parameters.
Therefore, it would not be easy to obtain a unified view of flares based on this formula.

%\citet{2011SSRv..159..263H}

We can expect from Fig. \ref{fig:EM_T} that the following assumptions of Shibata \& Yokoyama's scaling law hold even for small flares in the QS.
%\begin{itemize}
%\begin{description}
\begin{enumerate}[(I)]
\item Cooling by thermal conduction and heating by magnetic reconnection are balanced.
%Thus, the following equation is valid.
\begin{equation}
\kappa_0 \frac{T_{\mathrm{max}}^{7/2}}{2L^2} \simeq \frac{B_{\mathrm{corona}}^2}{4\pi}\frac{v_A}{L}, \label{cond_reco}
\end{equation}
where $\kappa_0\simeq 10^{-6}\,\mathrm{cgs}$ is the Spitzer thermal conductivity, $T_{\mathrm{max}}$ is the maximum temperature in the flare, and $v_{A}$ is the Alfv\'{e}n velocity.
\item The maximum flare temperature $T_{\mathrm{max}}$ is three times higher than the temperature observed at the flare peak time $T_{\mathrm{peak}}$.
\item The high-temperature, high-density plasma originating from the chromosphere is the origin of the brightening in the corona.
\begin{equation}
EM_v \simeq n^2L^3, \label{EMv_n}
\end{equation}
where $n$ is the increased flare-loop density.
\item The magnetic pressure of the loop confines the high-temperature, high-density plasma originating from the chromosphere.
\begin{equation}
2nk_{B}T_{\mathrm{peak}} \simeq \frac{B_{\mathrm{corona}}^2}{8\pi}, \label{pressure}
\end{equation}
where $k_{B}$ is the Boltzmann constant.
\end{enumerate}
%\end{description}
The fact that small flares in the QS also show these properties is essential for understanding the physical mechanism of nanoflares.

%These assumptions can also be confirmed by calculating physical quantities directly from the our observational results.
%The typical cooling rate by the thermal conduction $Q_{\mathrm{cond}}$ (in units of $\mathrm{erg}\,\mathrm{cm}^{-3}\,\mathrm{s}^{-1}$) calculated from the average of the parameters obtained in this study as follows.
%\begin{eqnarray}
%Q_{\mathrm{cond}} &=& \kappa_0 \frac{T_{\mathrm{peak}}^{7/2}}{2L^2} \notag \\
%&=&3.0\times 10^{-2} \left(\frac{T_{\mathrm{peak}}}{10^{6.15} \,\mathrm{K}} \right)^{7/2}\left(\frac{L}{2.5\times10^{8}\,\mathrm{cm}} \right)^{-2}. \label{ther_cond_obs}
%\end{eqnarray}
%optically thin cooling $Q_{\mathrm{rad}}$ (in units of $\mathrm{erg}\,\mathrm{cm}^{-3}\,\mathrm{s}^{-1}$)
%\begin{eqnarray}
%Q_{\mathrm{rad}} &=& n^2\Lambda(T_{\mathrm{peak}}) =8.0\times 10^{-4} \left(\frac{n}{2\times10^{9} \,\mathrm{cm}^{-3}} \right)^{2}. \label{rad_obs}
%\end{eqnarray}
%
%
%The possibility of the energy mechanism difference?

\subsection{Time evolution of flare temperature and density}     \label{sec4_3}

\begin{figure}
	\includegraphics[width=\columnwidth]{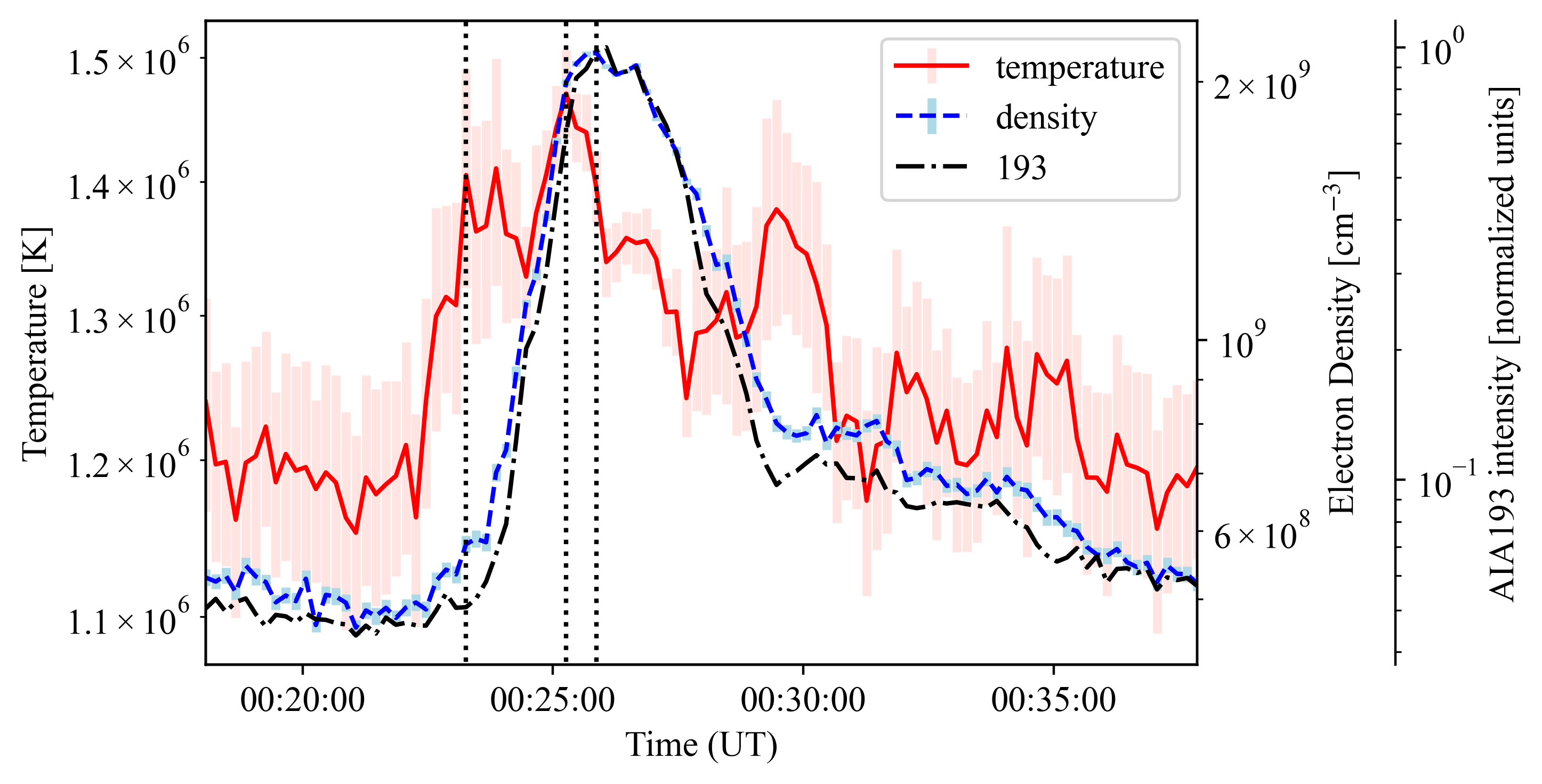}
    \caption{Typical example of the time evolution of density and temperature of a small flare.
         The event is the same as in Fig. \ref{fig:test}.
         The dotted black vertical lines indicate the times in Figs. \ref{fig:test}e, i, and m, respectively.
         The solid red, dashed blue, and dash-dotted black lines show the time evolution of the temperature, density, and 193 \AA\, intensity, respectively.
         Error bars of the temperature and density calculations are from the errors in estimating the DEM.
    }
    \label{fig:T_N_time}
\end{figure}

We investigated the time evolutions of temperature and density to understand how heating from small reconnection in the QS affects the atmosphere. 
The time evolution of flare temperature and density has been well studied in large solar flares by numerical simulations and observations \citep[e.g.,][]{1980SoPh...68..351N,1991A&A...241..197S,1992A&A...253..269J,1993A&A...267..586S,2002ApJ...577..422S}.
\citet{2007A&A...471..271R,2014LRSP...11....4R} summarised the time evolution of flare temperatures and densities in the following four phases \citep[see][fig.1. and 2.]{2007A&A...471..271R}.
\begin{enumerate}[\text{Phase} I:]
\item From the start of the heat pulse to the temperature peak (\textit{heating}).
\item From the temperature peak to the end of the heat pulse (\textit{evaporation}).
\item From the end of the heat pulse to the density peak (\textit{conductive cooling}).
\item From the density peak afterwards (\textit{radiative cooling}).
\end{enumerate}

We averaged the temperature and density at $3 \times 3$ pixels to investigate the time evolution.
To select pixels for the time evolution of temperature and density, we referred to the time when the AIA 193 \AA\, light curve was at its maximum.
At that time, we defined the $3 \times 3$ pixels, centred on the pixel with the largest AIA 193 \AA\, intensity.
We have fixed and limited the analysed pixels to facilitate comparison with the time evolution of the single-loop model \citep[e.g.,][]{1992A&A...253..269J}.

Fig. \ref{fig:T_N_time} shows a typical example of the time evolution of the density and temperature.
We can see from Fig. \ref{fig:T_N_time} that the temperature increase precedes the density increase.
The density and AIA 193 \AA\, light curves reach their peaks almost simultaneously after the temperature peaks. 
After the density peak, it decays a little more slowly than the AIA 193 \AA\, light curve.
The qualitative characteristics of these time evolutions are the same as those of large-scale flares \citep{2007A&A...471..271R,2014LRSP...11....4R}. 
The result that the temperature increase precedes the density increase suggests that chromospheric evaporation occurs even in small flares in the QS.
We found 15 events in which the temperature peak preceded the density peak.
This trend is consistent with the analysis of \citet{1998SoPh..182..349B}.

We note that the density increase simultaneously with coronal intensity increase was also found by \citet{2011A&A...529A..21K}, who described it as supporting chromospheric evaporation models.
However, only the fact that the density increase is synchronised with the coronal brightening is not enough to indicate the presence of chromospheric evaporation.
This is because optically thin coronal plasmas are sensitive to density fluctuations.
Moreover, numerical simulations suggest that the shocks also cause fluctuations in physical quantities comparable to those of nanoflares \citep{2004ApJ...601L.107M,2008ApJ...688..669A}.
%The preceding temperature than the density increase more strongly supports the scenario of energy release that occurs
%in the corona and propagates to the chromosphere, resulting in evaporation. 
The preceding temperature increase than the density increase more strongly supports the scenario that reconnection occurs in the corona and the released energy propagates to the chromosphere, resulting in evaporation.

\begin{figure*}
 \begin{center}
	\includegraphics[width=2\columnwidth]{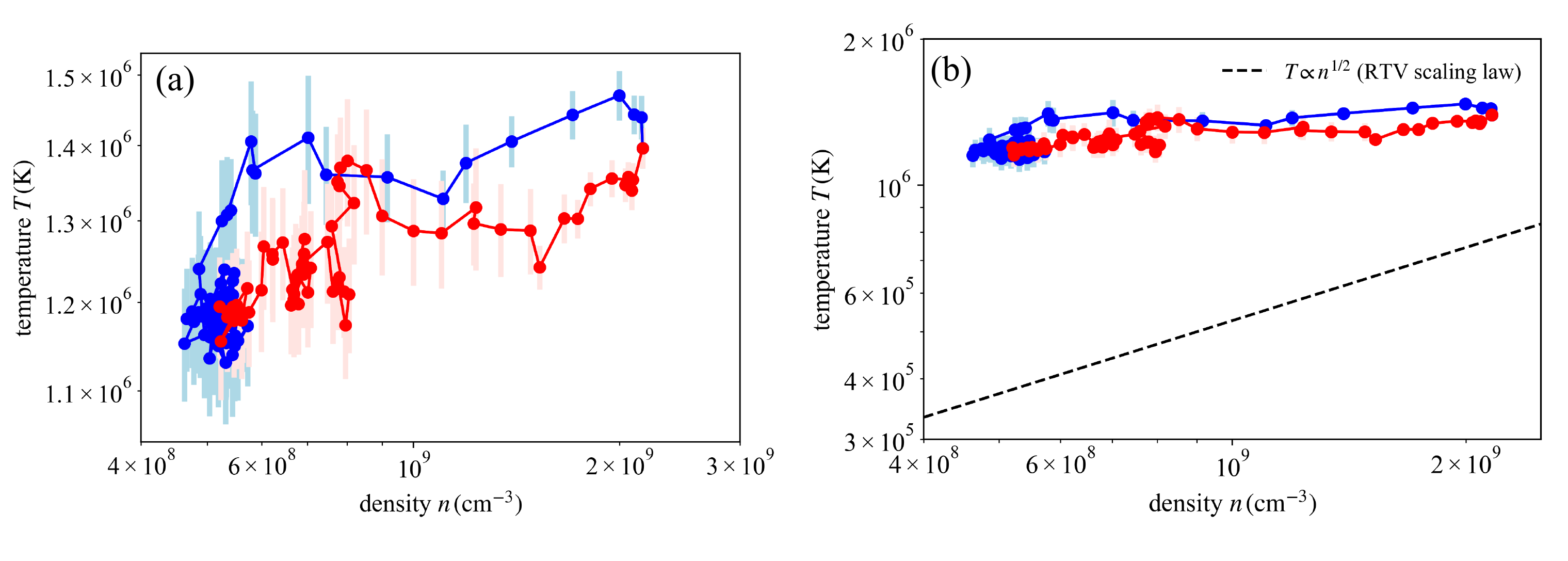}
    \caption{Typical example of the temperature-density (T-N) diagram.
         The event is the same as in Fig. \ref{fig:test}.
         (a): T-N diagram shown in large size. 
         (b): Comparison with RTV scaling law \citep[equation  \ref{RTV}, ][]{1978ApJ...220..643R}.
         The dashed black line is the RTV scaling law calculated at the spatial scale of event 4 in Table \ref{tab:example_table}.
         }
    \label{fig:T_N_diagram}
     \end{center}
\end{figure*}

Fig. \ref{fig:T_N_diagram} shows a typical temperature-density (T-N) diagram for small flares in the QS.
From Fig. \ref{fig:T_N_diagram}a, we can see that the temperature and density increase process is qualitatively the same as for large flares \citep{2007A&A...471..271R,2014LRSP...11....4R}. 
In contrast, Fig. \ref{fig:T_N_diagram}b shows that the temperature is about three times higher than that determined by the RTV scaling law \citep{1978ApJ...220..643R} in the pre and post-flare phases.
The power-law index of the decay phase is gradual and almost constant.

%The evolution shape itself is similar to that of a large-scale flare up to the peak of density.
%However, in the decay phase, the temperature decreases slowly without falling below the pre-flare temperature.
%The physical reason for this is expected to be that radiative cooling does not predominate as density increases.
%In other words, unlike a large flare, the density increase does not contribute to the temperature decrease; the temperature decreases by thermal conduction once the flare heating is finished.
%Density decreases slowly as it falls due to the gravity independent of temperature change.
%We can see from the figure that the entire time evolution follows a higher temperature than the RTV scaling law \citep{1978ApJ...220..643R}.

%Note that the reason the temperature did not drop below the pre-flare temperature during the decay phase may be that the DEM analysis by AIA six channels is almost insensitive to the chromosphere/transition region temperature.

\begin{figure}
	\includegraphics[width=\columnwidth]{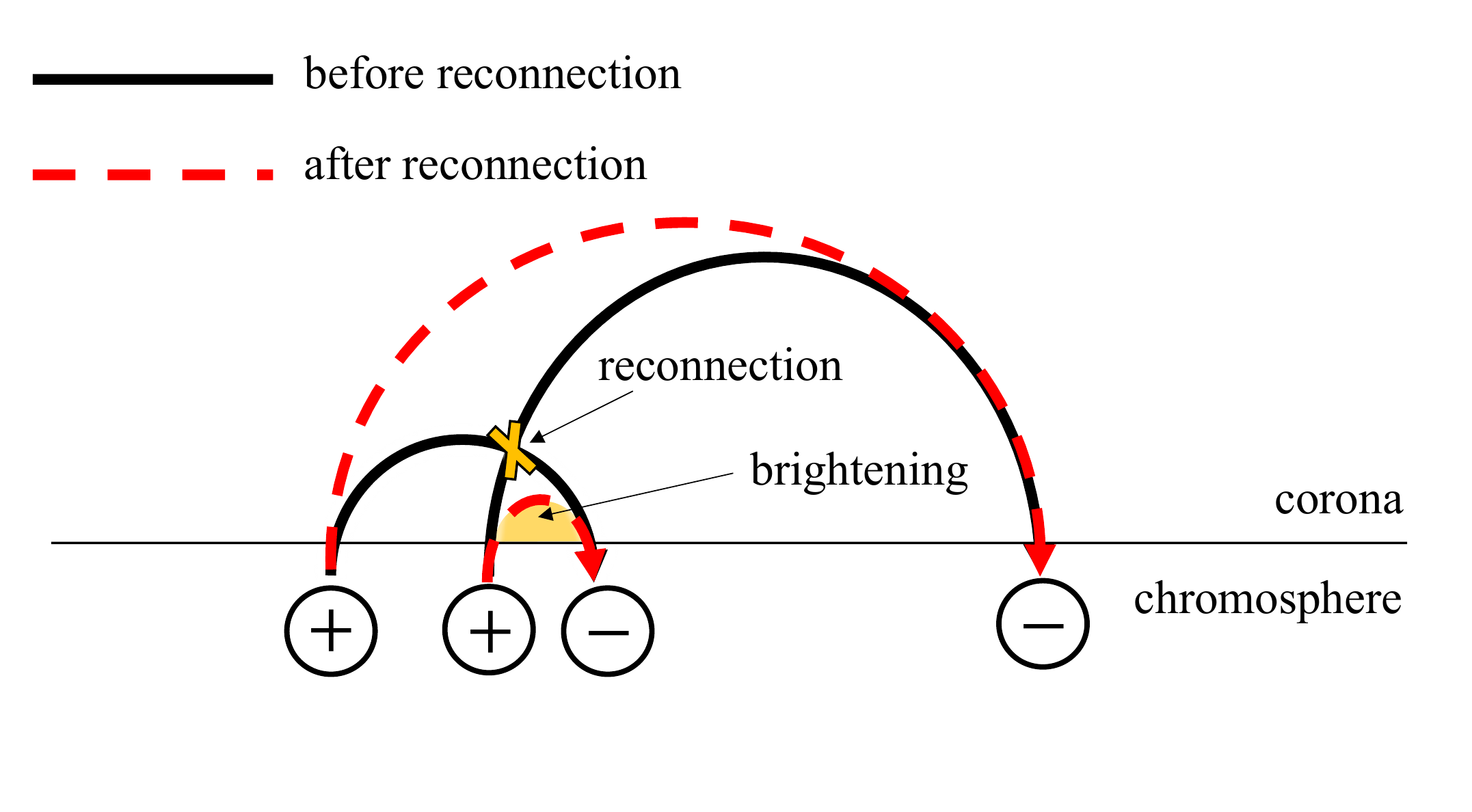}
    \caption{Schematic diagram showing an example of a magnetic field morphology that can underestimate the spatial scale.}
    \label{fig:imagine}
\end{figure}

There are multiple reasons for the higher temperatures than expected from the RTV scaling law, even without the occurrence of small flares.
The first is the possibility that small loops in the QS are not in a steady state.
The second possibility is that the spatial scale determined from the brightening pixels underestimates the actual loop length since brightening is seen only in part of the loop. %\citep{2021A&A...656L...4B,2021A&A...656A..35Z}.
%In fact,  \citet{2021A&A...656L...4B} proposed the possibility of brightening only near the top of the loop in small flares in the quiet region.
These underestimations will appear as a brightening only near the top of the loop \citep{2021A&A...656L...4B,2021A&A...656A..35Z}.
Also, if the brightening of a small loop formed by reconnection is observed primarily, as in Fig. \ref{fig:imagine}, the spatial scale will be underestimated.
Then the question to ask is: how much underestimation of the actual loop length would make the temperature consistent with the RTV scaling law when the flare is not occurring?
The RTV scaling law for the relationship between temperature and density is as follows:
\begin{equation}
T = 4.3\times10^5\left(\frac{n}{10^9\,\mathrm{cm}^{-3}}\right)^{1/2} \left(\frac{L}{2.5\times10^8\,\mathrm{cm}}\right)^{1/2}.     \label{RTV}
\end{equation}
Therefore, the spatial scale must be underestimated by approximately one order of magnitude.

A third possibility is that the DEM analysis using AIA itself was problematic.
The temperature response functions of the six AIA channels used in the DEM analysis have no peak around $10^{5.5}\sim3.2\times10^5\,\mathrm{K}$ \citep{2013ApJ...763...86L}.
The smallest temperature peak is about $10^{5.8}\sim6.3\times10^5\,\mathrm{K}$ in AIA 131 \AA, which is higher than the temperature of $4.3\times10^5\sim10^{5.63}\,\mathrm{K}$ expected from the RTV.
Therefore, the AIA DEM analysis may not be able to diagnose plasmas that is at a temperature as low as $10^{5.5}\,\mathrm{K}$.
We believe that the AIA DEM analysis may be the most influential in the above three possibilities since the temperature in the absence of flares is about $10^{6}\,\mathrm{K}$ for all events. 
Hence, the temperatures when flares are not occurring and during the decay phase will need to be investigated differently.

%Someone might think that the problem with the temperature diagnostics by the DEM analysis mentioned above applies not only to the pre-flare or decay phase but also to the flare peak time, shown in Fig. \ref{fig:EM_T}.
%However, we expect the results of DEM analysis by AIA to be reliable near the flare peak time.
%If the main component of the flare is low-temperature ($\sim10^5\,\mathrm{K}$) plasma that cannot be captured by DEM analysis, magnetic reconnection will occur in or below the transition region.
%This is because, considering the AIA temperature response function, low-temperature plasmas must be high-density ($\sim10^{10}\,\mathrm{cm^{-3}}$) to explain the small brightenings \citep{2023arXiv230102040D}.
%However, we found evidence from time variations in the physical quantities (Fig. \ref{fig:T_N_time}) and light curves (Fig. \ref{fig:light_curve}) that reconnection occurred above the transition region in more than half of the events.
%Thus, most of the events analysed in this study are expected to have temperatures above $\sim10^6\,\mathrm{K}$ around the flare peak time.
Although one may speculate that the temperature diagnostics problem identified by the DEM analysis above could extend to the flare peak time, as illustrated in Fig. \ref{fig:EM_T}, we anticipate that the results of DEM analysis by AIA are reliable in the vicinity of the flare peak time.
If the primary component of the flare is low-temperature ($\sim10^5\,\mathrm{K}$) plasma that is not detectable by DEM analysis, magnetic reconnection is likely to occur in or below the transition region.
This is because, according to the AIA temperature response function, low-temperature plasmas must have a high density ($\sim10^{10}\,\mathrm{cm^{-3}}$) to account for the observed small brightenings \citep{2023A&A...671A..64D}.
However, our investigation revealed evidence of reconnection occurring above the transition region in more than half of the events, as demonstrated by the temporal variations in physical quantities  (Fig. \ref{fig:T_N_time}) and light curves (Fig. \ref{fig:light_curve}).
Thus, we expect most of the events examined in this study to have temperatures exceeding ($\sim10^5\,\mathrm{K}$) around the flare peak time.

\begin{figure*}
	\includegraphics[width=2\columnwidth]{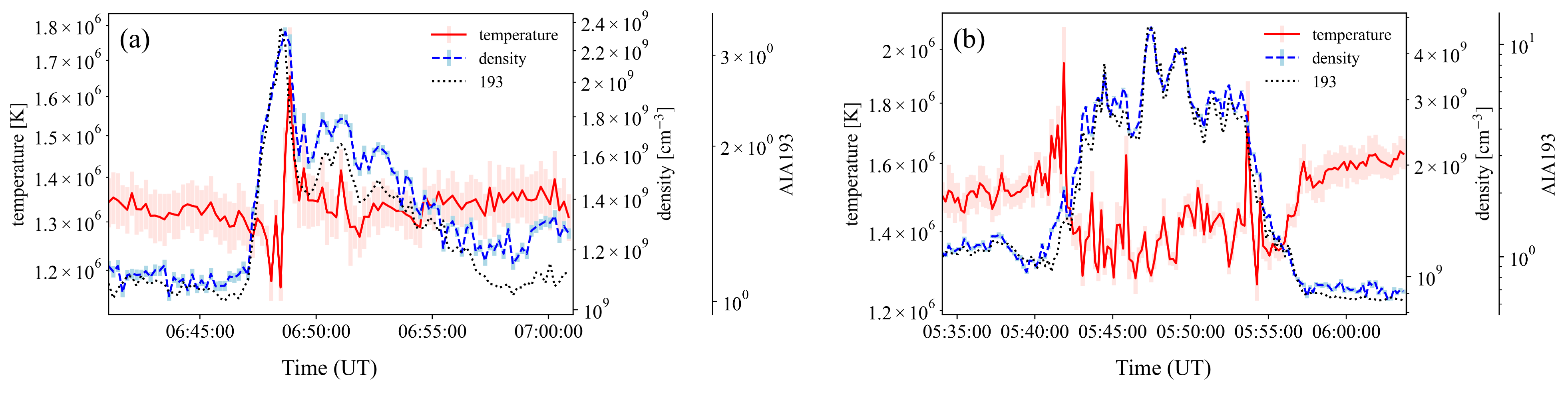}
    \caption{Examples of the temperature and density evolution that differ from typical events.}
    \label{fig:strange}
\end{figure*}

We present two examples of temperature and density temporal evolution showing peculiar behaviour. 
The first example is events in which the density peak precedes the temperature peak (Fig. \ref{fig:strange}a).
We identified five such examples.
One of the causes of these events is the failure to fully resolve the single loop. 
In other words, the temperature increase corresponding to the density peak may not have been captured by the selected pixels. 
In such cases, we may capture the temperature increase corresponding to the small brightenings that repeatedly occur after the density peak. 
Another possibility is that reconnection below the transition region heated the dense plasma to coronal temperatures \citep{2021A&A...656L...7C}.
Since the AIA DEM analysis basically describes plasma at coronal temperatures, the observed EM value should increase when plasma at chromospheric/transition region temperatures is heated to coronal temperatures. 
In such cases, the plasma in the chromosphere/transition region should be heated to coronal temperatures with decreasing density due to expansion. 
Thus, the observed temperature should peak at or slightly after the density peak. 

The second example is that the temperature decreases at the density peak with respect to the pre-flare one (Fig. \ref{fig:strange}b).
We identified four such examples.
These events occurred at CBPs and had relatively high temperatures ($T\sim1.5\times10^{6}\,\mathrm{K}$) before the flares occurred.
Therefore, the low temperature ($T\sim1.2\times10^{6}\,\mathrm{K}$) plasma is considered to have increased due to chromospheric evaporation, resulting in a decrease in temperature at the density peak. 
We need to test this hypothesis in the future, including the perspective of what determines the temperature of chromospheric evaporation.

\section{SUMMARY AND FUTURE WORK}

This paper presents our analysis of many small flares in the QS using SDO/AIA and SMART/SDDI observational data, with particular attention to their thermal properties.
Our analysis confirmed that the events had physical quantities that were quantitatively consistent with previous studies.
We also confirmed common qualitative properties, such as that they occurred at network boundaries and that most of them were accompanied by flux cancellation. 
The main results of this paper are as follows:
\begin{enumerate}[(i)]
\item 
Redshift with brightening of the H$\alpha$ line centre was observed in more than half of the events. 
These redshifts corresponded well to brightenings in the corona both temporally and spatially and were also consistent with the chromospheric condensation formula \citep{2014ApJ...795...10L}.
Therefore, they are considered to correspond to chromospheric condensation for small flares in the QS (Sections \ref{sec3_2}, \ref{sec3_3}, and \ref{sec4_1}).
\item 
The observed relationship between EM and temperature is consistent with the scaling law \citep{1999ApJ...526L..49S,2002ApJ...577..422S} for coronal magnetic field strengths of 5--15 G.
This consistency suggests that the primary cooling mechanism at the flare peak time is thermal conduction and that the plasma originated from the chromosphere through evaporation significantly affects the brightening.
In other words, it suggests that small flares in the QS also have qualitatively the same thermal properties in their peak time, only with reduced magnetic field and spatial scale (Section \ref{sec4_2}).
\item
In more than half of the events, the temperature reached a maximum before the density.
This result supports the idea that chromospheric evaporation occurs even in some small flares in the QS (Section \ref{sec4_3}).
\item 
Our temperature-density diagram shows that the thermal evolution of small flares always proceeds at higher temperatures than the RTV scaling law \citep{1978ApJ...220..643R}.
One of the reasons for this evolution at high temperatures is the possibility that we overestimate the temperatures in the steady state and decay phase due to the limitation of the observed temperature range by AIA (Section \ref{sec4_3}).
\end{enumerate}
These results suggest that the interaction with the chromosphere may play an essential role in the thermal evolution of some small flares in the QS.
This property may be the new commonality between small flares in the QS and typical active region flares.
However, due to the lack of sample size and spatial resolution, our observations do not accurately answer the question of how many small flares in the QS interact with the chromosphere.
Our observations also may not have captured the entire thermal evolution of the small flare in the QS.
To solve these problems, simultaneous higher spatial resolution spectroscopic observations of the chromosphere  (e.g., with SST/CRISP or DKIST/VTF) and observations capable of diagnosing temperatures from $10^4$ to $10^6\,\mathrm{K}$ with equally high spatial and temporal resolution (e.g., IRIS or Solar-C/EUVST) will be necessary.

We propose some suggestions based on our results in determining the relationship between the frequency and energy of small flares in the QS. 
To determine the power-law index of the relationship, we need a proper detection method and method for determining flare energy over a wide energy range.
%Regarding the flare detection method, our study suggests that using coronal intensity or EM variations may not be able to detect events that only produce heating without a brightening.
%This is because the density and coronal intensity showed almost the same time evolution in our analysis, whilst the temperature showed a behaviour independent of them. 
%In other words, because coronal intensity and temperature can be determined independently, we cannot rule out the existence of reconnection, in which the density hardly increases relative to the temperature.
%This hypothesis would correspond to a QS version of the heating events without X-ray brightening observed in active regions \citep{2017NatAs...1..771I}.
%The first candidates for such reconnection are small ones in braided field lines \citep{1988ApJ...330..474P,2021NatAs...5...54A} or weak magnetic fields outside the network boundary.
%Another candidate, if reconnection occurs in a loop that is already reconnected and highly dense, with radiation cooling as the primary cooling mechanism, would also not cause significant intensity fluctuations.
%If these events were to occur, how to detect heating (reconnection) events would need to be carefully verified, including numerical verification.
Regarding the flare detection method, our study suggests that using coronal intensity and EM fluctuations as detection criteria might be ineffective in identifying heating events with slight intensity variations.
This is because the density and intensity of the corona showed similar time evolution in our analysis, but the temperature behaved differently from them (Figs \ref{fig:T_N_time} and \ref{fig:strange}).
In other words, density rather than temperature provides the major contribution to the coronal intensity.
These heating events with slight intensity variations would correspond to QS versions of the heating events without X-ray brightening observed in active regions \citep{2017NatAs...1..771I}.
%Candidates for heating with slight density variation would be cases where reconnection occurs but the energy flowing into the chromosphere is too small for chromospheric evaporation to occur.
Potential instances of heating events characterised by minimal density fluctuations could include scenarios where magnetic reconnection occurs, yet the energy transported into the chromosphere remains insufficient to trigger chromospheric evaporation.
Such circumstances may arise when the magnetic field strength is exceptionally weak or when reconnection occurs on tiny spatial scales, as in braiding \citep{1988ApJ...330..474P,2021NatAs...5...54A}.
%In addition, density variations would be small if reconnection occurred in a high-density loop with evaporation already occurring.
Moreover, density variations might be negligible if reconnection occurs within a high-density loop already experiencing evaporation.
This situation can be rephrased as one where the reconnection cadence is shorter than the plasma cooling time \citep{2016A&A...591A.148J}.
If these events were to occur, how to detect heating (reconnection) events would need to be carefully verified, including numerical verification.

To determine the flare energy, we propose estimating the radiation energy in the chromosphere of a small flare in the QS.
The energy partition of flares, including large flares, is an open question.
In particular, \citet{2020A&A...644A.172W} has proposed that  thermal-nonthermal energy partition changes with flare energy.
Hence, using thermal energy as the flare energy may be an inappropriate definition for comparing flares with a wide range of energies.
On the other hand, the bolometric radiated energy is considered a good proxy for the dissipated magnetic energy of the flare. 
The bolometric energy is estimated from variations in the total solar irradiance (TSI) \citep{2011A&A...530A..84K,2012ApJ...759...71E}, with the main contribution coming from visible and UV wavelengths.
It would be impossible to estimate the bolometric energy from TSI variations for small flares in the QS; however, it is possible to estimate the radiative energy of the chromosphere from non-LTE inversion \citep{2022A&A...665A..50Y}.
By estimating the radiative energy of the chromosphere for small flares, the flare energy can be defined more accurately, which may help solve the coronal heating problem.

\section*{Acknowledgements}

%We thank for fruitful discussions and comments.
We thank M. Madjarska for fruitful discussions and checking this manuscript.
We also thank Y. Chen for fruitful discussions.
We wish to thank the anonymous referee for helpful comments that led to improvements in this work.
We are grateful to the staff of Hida Observatory for the instrument development and daily observations.
We are grateful to the SDO/AIA teams. 
SDO is part of NASA’s Living with a Star Program. 
%We want to thank Editage (www.editage.com) for English language editing.
This research is supported by JSPS KAKENHI grant numbers 22J14637 (Y.K.), 21H01131 (K.S., K.I., and A.A.), and 21J14036 (D.Y.).

\section*{DATA AVAILABILITY}

The SDO/AIA and SDO/HMI data underlying this article are available in \url{http://jsoc.stanford.edu/ajax/lookdata.html}.
SMART/SDDI data are available at \url{https://www.hida.kyoto-u.ac.jp/SMART/daily/19Sep/daily/images_20190907.html} for some wavelengths.
All wavelength data for SMART/SDDI can be obtained by requesting data\_info@kwasan.kyoto-u.ac.jp.
The derived data generated in this work will be shared on reasonable request to the corresponding author.

%%%%%%%%%%%%%%%%%%%%%%%%%%%%%%%%%%%%%%%%%%%%%%%%%%



\bibliographystyle{mnras}
\bibliography{draft_bb_full.bib} % if your bibtex file is called example.bib

%\begin{thebibliography}{99}
%\end{thebibliography}

%%%%%%%%%%%%%%%%%%%%%%%%%%%%%%%%%%%%%%%%%%%%%%%%%%

%%%%%%%%%%%%%%%%% APPENDICES %%%%%%%%%%%%%%%%%%%%%

\appendix

%\section{The power-law index between temperature and density in the decay phase}

%%%%%%%%%%%%%%%%%%%%%%%%%%%%%%%%%%%%%%%%%%%%%%%%%%
%Figures and tables should be placed at logical positions in the text. Don't
%worry about the exact layout, which will be handled by the publishers.
%
%Figures are referred to as e.g. Fig.~\ref{fig:example_figure}, and tables as
%e.g. Table~\ref{tab:example_table}.

%\begin{figure}
%	% To include a figure from a file named example.*
%	% Allowable file formats are eps or ps if compiling using latex
%	% or pdf, png, jpg if compiling using pdflatex
%	\includegraphics[width=\columnwidth]{example}
%    \caption{This is an example figure. Captions appear below each figure.
%	Give enough detail for the reader to understand what they're looking at,
%	but leave detailed discussion to the main body of the text.}
%    \label{fig:example_figure}
%\end{figure}

%\begin{table}
%	\centering
%	\caption{This is an example table. Captions appear above each table.
%	Remember to define the quantities, symbols and units used.}
%	\label{tab:example_table}
%	\begin{tabular}{lccr} % four columns, alignment for each
%		\hline
%		A & B & C & D\\
%		\hline
%		1 & 2 & 3 & 4\\
%		2 & 4 & 6 & 8\\
%		3 & 5 & 7 & 9\\
%		\hline
%	\end{tabular}
%\end{table}

% Don't change these lines
\bsp	% typesetting comment
\label{lastpage}
\end{document}